# Electronic Transport and quantum oscillation of Topological Semimetals

Jin Hu[1], Su-Yang Xu[2,3], Ni Ni[4], Zhiqiang Mao[5]

[1]Department of Physics and Institute for Nanoscience and Engineering, University of Arkansas, Fayetteville, AR 72703, USA

[2]Department of Physics, Massachusetts Institute of Technology, Cambridge, Massachusetts 02139, USA

[3]Laboratory for Topological Quantum Matter and Spectroscopy (B7), Department of Physics, Princeton University, Princeton, New Jersey 08544, USA

[4]Department of Physics and Astronomy and California NanoSystems Institute, University of California, Los Angeles, CA 90095, USA

[5]Department of Physics, Pennsylvania State University, University Park, PA 16802, USA

## Abstract

Discoveries of three-dimensional (3D) topological semimetals have generated tremendous excitement in the scientific community, since they represent a new class of topological matters and carry great promise for technological applications. The study of this family of materials has been at the frontiers of condensed matter physics and many breakthroughs have been made. Up to date, several topological semimetal phases, including Dirac semimetals (DSMs), Weyl semimetals (WSMs), nodal-line semimetals (NLSMs) and triple-point semimetals, have been theoretically predicted and experimentally demonstrated by angle-resolved photoemission experiments. The low energy excitation around the Dirac/Weyl nodal points, nodal line, or triply degenerated nodal point can be viewed as emergent relativistic fermions. Ideal relativistic fermions all show distinct properties, such as zero mass, extremely high mobility, and Berry phase in cyclotron motion. In addition, relativistic fermions in 3D topological semimetals have also been found to exhibit a rich variety of exotic transport properties, e.g. extremely large magneto-resistance in DSMs/WSMs/NLSMs, chiral anomaly and in-plane Hall effect in WSMs, intrinsic anomalous Hall effect in time reversal symmetry breaking WSMs, quantum oscillations due to Weyl orbits in DSMs, quantum tunneling of the relativistic fermions of the zeroth-Landau level in layered DSMs, etc. Furthermore, quantum spin Hall and quantum anomalous Hall effects are predicted to occur when the dimensionality of some topological semimetals is reduced from 3D to 2D. Both quantum spin Hall and quantum anomalous Hall effects originate from topological edge states and can support dissipationless transport, which are of great potential for applications in electronics and spintronics. In this review article, we first briefly introduce band structural characteristics of each established topological semimetal phase; then we are focused on two topics: electronic transport and quantum oscillations of topological semimetals. Since topological semimetals are commonly characterized by high carrier mobility, they often show strong quantum oscillations even at high temperatures and low/moderate field ranges. We first review quantum oscillation studies of topological semimetals and give detailed discussions on how to extract relativistic fermion properties from the analyses of Shubnikov-de Haas (SdH) and De Haas-van Alphen (dHvA) quantum oscillation data. Then we review the current studies on exotic transport properties mentioned above and discuss how they are connected to non-trivial band topology and make comment on the perspective of this area at the end.

1. **Introduction**

The rich cross-pollination between high energy and condensed matter physics has led to deeper knowledge of important topics in physics such as spontaneous symmetry breaking, phase transitions and renormalization (1; 2). Such knowledge has, in turn, greatly helped physicists and materials scientists to better understand magnets, superconductors and other novel materials, leading to practical device applications (1). In the past decade, there has been significant interest in realizing high energy particles in solid state systems. The theoretical attempts to explain graphene's properties (3) using solid-state physics led to an equation similar to one otherwise seen in cosmology and colliders: the Dirac equation. Following graphene's discovery, many materials with nodal band crossings, known as topological insulators and semimetals (4-11), were discovered, generating significant research excitement. The topological Dirac (12-14) and Weyl (2; 15-23) semimetals are crystalline solids whose low-energy electronic excitations resemble the Dirac (24) and Weyl (15) fermions in high-energy particle physics, respectively. In particular, although the Weyl fermion played a crucial rule in the Standard Model (15), it has never been observed as a fundamental particle. The realization of the topological Weyl semimetal state (22; 23; 25-27) enables the observation of this elusive particle in physics. Topological semimetals further allow for band crossings beyond high-energy classifications. Primary examples include the type-II Weyl (28) and Dirac (29) semimetals, the nodal-line semimetals (30), and the unconventional fermion semimetals (31-36). Due to the rich variety of crystalline and magnetic symmetry properties of condensed matter systems (37), it is likely that this is only the tip of an iceberg and there are ample new topological semimetals awaiting discovery. These topological semimetals provide platforms for studying a number of important concepts in high energy physics (e.g. the chiral anomaly) in table-top experiments. Moreover, they extend the classification of topological phases from gapped matter (e.g. insulators) to gapless systems (e.g. metals).

Topological semimetals enable a kaleidoscope of novel electronic properties. They support exotic, topologically protected boundary modes such as the topological Fermi arcs and drumhead surface states. These surface states have been directly observed in spectroscopic measurements (19; 25; 27; 38-42). The Fermi arcs also lead to unusual quantum cyclotron orbits (the Weyl orbits) as observed in quantum oscillation measurements (43; 44). Because of the linear dispersion and the spin (pseudo-spin) momentum locking, low-energy electrons in topological semimetals are highly robust against crystalline disorder and imperfections, leading to very high electron mobilities (45; 46). The compensating electron and hole carriers further cause non-saturating magnetoresistance (46-48) and magneto-thermopower (49-51). The application of parallel electric and magnetic fields can break the apparent conservation of the chiral charge (10; 11; 52; 53). Such chiral anomaly leads to enhanced conductivity with an increasing magnetic field. The diverging Berry curvatures near the nodal points support distinct anomalous transport phenomena including intrinsic anomalous Hall effects (54-56), and anomalous Nernst effects (57; 58). They also support significantly enhanced optical and optoelectronic phenomena including large (even quantized) photocurrents (59-64), second-harmonic generation (65; 66), optical activity and gyrotopy (67-69) and Kerr rotation (70; 71). Furthermore, thinning down a three-dimensional (3D) topological semimetal into two-dimensions (2D) may give rise to new 2D topology including the quantum spin Hall insulator and the quantum anomalous Hall insulator (14; 20; 21; 72-76). These unconventional transport and optical properties of topological semimetals can pave the way for the realization of dissipationless electronic and spintronic devices as well as efficient photodetectors and energy harvesters.

The area of 3D topological semimetals is fast growing; many papers have been published on theoretical predictions and experimental studies. There have been many review articles, which introduce progresses in theoretical and experimental studies on topological semimetals (8-11; 76-85). In this review article, we are focused on electronic transport and quantum oscillation studies on topological semimetals; these two topics have not been reviewed comprehensively in previous review articles. Before we start detailed discussions on these topics, we will first briefly introduce each prototype topological semimetal phase and discuss their band structure characteristics, topological invariant and symmetry protections.

## 2. Categories of topological semimetals

In this section, we discuss various three-dimensional topological semimetal phases of matter, including Weyl semimetals, Dirac semimetals, nodal-line semimetals, and unconventional fermion semimetals beyond the Dirac and Weyl paradigm. For each kind of topological semimetal, we focus on three aspects, the characteristic band structure (the number of bands that cross, the dimensionality of the band crossing in $k$ space, and the typical energy-momentum dispersion), the topological invariant and the symmetry protections, and representative materials.

## 2.1 Weyl semimetals

Weyl semimetals are a class of topological semimetals that host Weyl fermions as low energy quasiparticle excitations (2; 6-11; 15-21). In a Weyl semimetal, two singly degenerate bands cross at discrete points, i.e. Weyl nodes, and disperse linearly in all three momentum space directions away from each Weyl node (Fig. 1a). Weyl fermions have distinct chiralities, either left-handed or right-handed. The chiralities of the Weyl nodes give rise to chiral charges, which can be understood as monopoles and anti-monopoles of Berry flux in momentum space. The separation of the opposite chiral charges in momentum space leads to surface Fermi arcs, whose constant energy contours are open arcs that connect the Weyl nodes of opposite chiralities on the surface.

Because of the existence of Weyl nodes, Weyl semimetals lack a global band gap. The absence of a global band gap prevents the definition of a topological invariant for the entire three-dimensional (3D) bulk Brillouin zone (BZ). On the other hand, on a two-dimensional (2D) closed surface that encloses the Weyl node in $k$ space (Fig. 1a), the band structure is fully gapped and therefore allows for a topological invariant to be defined (19). Specifically, the Chern number associated with the 2D closed surface directly corresponds to the topological invariant of a Weyl node (i.e., the chiral charge). Mathematically, the chiral charge $C$ can be calculated by the integral of the Berry curvature (the Berry flux) as shown below

$$C = \int_S \mathbf{\Omega} \cdot d\mathbf{S} \qquad (1)$$

where $S$ is the 2D closed surface in $k$ space that encloses the Weyl node and $\mathbf{\Omega}$ is the Berry curvature. Due to the chiral charge, Weyl nodes can appear at generic $k$ points of the BZ. In the presence of translational symmetry, these Weyl nodes are topologically stable and cannot be removed without pair-annihilation. The existence of Weyl nodes does not rely on any additional crystalline point group symmetries.

Real materials that host the Weyl semimetal (WSM) state are usually further classified as inversion-symmetry breaking WSM and time-reversal symmetry breaking WSM. Representative inversion symmetry breaking WSMs include the TaAs family of noncentrosymmetric crystals (22; 23; 25; 27; 39; 86-92). Representative time-reversal symmetry breaking WSMs can be realized in naturally occurring ferromagnetic semimetals such as pyrochlore irridate (19), $HgCr_2Se_4$ (21), $Co_3Sn_2S_2$ (93; 94), Heuslers (95-99), and the non-collinear antiferromagnets $Mn_3Sn$ and $Mn_3Ge$ (57; 100-103) or by applying an external magnetic field to a nonmagnetic or antiferromagnetic semimetal as demonstrated in the magneto-transport experiments (104) on $Na_3Bi$ (105), $Cd_3As_2$ (45; 106), $ZrTe_5$ (107) and half Heuslers (108-110). From a different angle, WSMs can also be classified by the energy-momentum dispersions near the Weyl nodes. Type-I Weyl semimetals have un-tilted or weakly tilted Weyl cones with point-like Fermi surface when the chemical potential is placed at the Weyl node. By contrast, type-II WSMs have strongly tilted Weyl cones (Fig. 1b) (28). Their Fermi surface consists of electron and hole pockets that touch at the type-II Weyl nodes. Representative type-II WSMs include $WTe_2$ (28; 111-113), $MoTe_2$ (114-122), $TaIrTe_4$ (123; 124) and $(W/Mo)P_2$ (125). We note that these different classifications are not mutually exclusive. For instance, $MoTe_2$ is not only an inversion breaking WSM, but also a type-II WSM.

## 2.2 Dirac semimetals

Dirac semimetals (DSMs) host Dirac fermions as low energy quasiparticle excitations (12-14; 38; 126-131). In a DSM, two doubly degenerate bands cross to form a Dirac node, and disperse linearly in all three momentum directions away from the node. Each Dirac node can be viewed as a pair of degenerate Weyl nodes of opposite chiralities. Since a pair of degenerate Weyl nodes of opposite chiralities is in general unstable and may annihilate, additional crystalline point group symmetries are needed to realize a stable DSM phase (131). One route is to rely on uniaxial rotational symmetries (131). Specifically, a band inversion can create a pair of 3D Dirac nodes on the opposite sites of the time-reversal invariant momenta.

Representative Dirac semimetals of this kind include Na$_3$Bi (13; 38; 126) and Cd$_3$As$_2$ (14; 127-130) (type-I) as well as VAl$_3$ (29) (type-II). Another route is to rely on non-symmorphic symmetries, i.e. glide reflections and screw rotations. It has been theoretically shown that non-symmorphic symmetries can lead to nontrivial band connectivity at the BZ boundaries, giving rise to filling-enforced DSMs or nodal-line semimetals depending on the specific space groups (12; 132-135). Representative filling-enforced DSM candidates include $\beta$-BiO$_2$ (12) and distorted spinels (132). Furthermore, a DSM can be realized as the critical point of the topological phase transition between a trivial insulator and a topological insulator. This is achieved in the BiTl(S$_{1-x}$Se$_x$)$_2$ (12; 136), Bi$_{2-x}$In$_x$Se$_3$ (137) and Pb$_{1-x}$Sn$_x$Te (138) systems by fine tuning the chemical doping concentration. Alternatively, compounds like ZrTe$_5$ (107; 139; 140) and the SrMnSb$_2$ (141-143) family naturally sit near the critical point of such a topological phase transition and therefore approximate a DSM state. According to current theoretical understanding, Dirac nodes are not associated with any nontrivial topological invariant (i.e. they have zero chiral charge) (144).

## 2.3 Nodal-line semimetals

In nodal-line semimetals (NLSM), conduction and valence bands cross at one-dimensional (1D) lines in $k$ space (Fig. 1a) (30; 40; 78; 85; 133; 134; 145-161). Compared to DSMs/WSMs, the electronic structure of NLSMs is distinct in three aspects: (1) The bulk Fermi surface consists of 1D lines in NLSMs but 0D points in WSMs; (2) The density of states is proportional to $(E - E_F)^2$ in NLSMs but $|E - E_F|$ in WSMs; (3) On the surface, nodal lines are accompanied by "drumhead"-like surface states, whereas Weyl nodes are connected by 1D Fermi arc surface states.

We now discuss the topological invariant of NLSMs. We consider a 1D closed loop that inter-links the nodal line in $k$ space (Fig. 1c). The band structure is fully gapped and therefore allows for the definition of a topological invariant, i.e., the winding number (150). Mathematically, the winding number $w$ is defined as the integral of the Berry connection along the 1D closed loop that links the nodal line as shown below

$$w = \int_l \mathbf{A} \cdot d\mathbf{l} \qquad (2)$$

where *l* is the 1D closed loop that links the nodal-line and **A** is the Berry connection.

NLSMs also come in a variety of forms depending on the characteristic band structure and the symmetry protection. First, nodal-lines can be closed loops inside the 3D BZ (also called nodal-circles). Such nodal-circles are naturally formed by a band inversion. The nodal-circles are further classified based on the symmetry protection. There are nodal-circles that are only strictly gapless in the absence of spin-orbit coupling (SOC) (78; 146; 149; 150). They are usually protected by the combination of time-reversal and inversion symmetries (78; 146; 150). Representative materials include $Cu_3N$ (149), $Ca_3P_2$ (147), $Cu_3PdN$ (148) and the ZrSiS family (154-158). Alternatively, nodal-circles can be formed in noncentrosymmetric crystals protected by a mirror plane. These nodal-circles are stable even upon the inclusion of SOC. Representative materials include $PbTaSe_2$, $TlTaSe_2$ and CaAgAs (40; 145; 159; 160). Second, nodal-lines can also be a straight line that span across the BZ. Representative materials include the $BaNbS_3$ family (161). Third, the nodal-circles can interlink with each other in *k* space forming Hopf links and nodal chains (162-167). They may be protected by the presence of multiple perpendicular mirror planes (167) or by nonsymmorphic symmetries (162; 163).

## 2.4 Unconventional fermion semimetals

In contrast to high energy physics, solid state crystals can support band crossings beyond the Dirac/Weyl paradigm (31-36). These band crossings, broadly referred as "unconventional fermions", include 3-, 4-, 6-, and 8-fold degeneracies (31).

Here we take a particular type of 3-band crossing as an example (33-36; 168-170). In such a triple-point semimetal, three singly degenerate bands cross at discrete points, the triple-points (Fig. 1d). Moving away from one triple-point along $k_x$ or $k_z$, the three bands all become non-degenerate. By contrast, moving away along $k_y$, bands 1 and 2 remain degenerate for $-k_y$ whereas bands 2 and 3 remain degenerate for $+k_y$. Therefore, the triple-point can also be viewed as the meeting point between two nodal lines along the $k_y$ axis. These triple-points are protected by the combination of a uniaxial rotational axis, mirror planes and time-reversal symmetry. These triple-points are not associated with any topological invariant due to the lack of a global band gap on any 2D closed surface that encloses the triple-point. Representative materials include MoC, WC, MoP, and ZrTe (33-36; 169; 170).

## 3. Transport signatures of topological semimetals

The relativistic nature of the Dirac and Weyl fermions in topological semimetals manifests in many distinct transport properties including extremely large magnetoresistance, high mobility, light effective mass, non-trivial Berry phase, Chiral anomaly and anomalous Hall effect. These relativistic fermion properties host great potential for future electronic and spintronic applications. Characterization of relativistic fermions through transport measurements provides a convenient approach to verify a non-trivial topological state, complementary to the direct observation of non-trivial band topology by ARPES experiments. In this section, we will summarize these transport signatures of topological DSMs and WSMs.

### 3.1 Magnetoresistance

The electron transport in topological semimetals is usually strongly affected by external magnetic field. Large magnetoresistance (MR) is a common signature often seen in most DSMs and WSMs. MR is usually expressed as the change of resistance (resistivity) under field normalized by the zero-field resistance (resistivity), i.e., $[R(B) - R(B=0)]/R(B=0)$ or $[\rho(B) - \rho(B=0)]/\rho(B=0)$. The transverse MR, measured with the field perpendicular to the current direction, can reach up to 0.1 - 1 million percent at low temperatures (0.5 – 5 K) and a field of 9 T (see Table 1), without any sign of saturation up to 30 – 100 T in WSMs/DSMs such as $Cd_3As_2$, $PtBi_2$, $WTe_2$, NbP, etc (46; 48; 171; 172). A power law field dependence (MR $\propto B^n$) is usually seen in various topological semimetals, with the exponent *n* ranging from 1 to 2 (45; 46; 48; 107; 171-187).

In a simple metal, a positive transverse MR with quadratic field dependence is generally expected due to the Lorentz effect (47). Such Lorentz effect-induced "orbital MR" is usually weak and saturates for systems with a closed Fermi surface, contrasted with giant, non-saturating MR seen in topological semimetals. The origin of the unusually large MR of topological semimetals has been intensively studied. Electron-hole compensation has been proposed to be a possible mechanism (46; 48; 171). However, there are also reports which indicate carrier compensation is not achieved in some topological semimetals (188; 189). An alternative explanation is that the backscattering at zero field is strongly suppressed by some protection mechanisms associated with non-trivial band topology, and significantly enhanced by magnetic fields (45).

The strong coupling between MR, high mobility, and linearly dispersed Dirac/Weyl cones may provide some clues for further understanding of the large MR. High mobility ($\mu$) is another signature accompanied with large MR in topological semimetals. Mobility is related with conductivity $\sigma$ via $\sigma = nq\mu$, where $n$ and $q$ are the carrier density and charge, respectively. For a single band system, the Hall coefficient $R_H = 1/nq$ and thus $\mu = \sigma \cdot R_H$. However, in multiple-band systems, the field dependence of Hall resistivity $\rho_{xy}$ deviates from linearity. Fig. 2a shows one example. In this case, the Hall coefficient, defined as $d\rho_{xy}/dB$, becomes field dependent and both mobility and carrier density cannot be directly derived as for a single band system. A commonly used approach for analyzing the transport properties of multiband systems is the multiple-band model, i.e, assuming that the contributions of various bands to the conductivity are additive. In practice, for a system with more than two bands, a further simplified model, which considers only one electron- and one hole-band, is widely used to describe the longitudinal resistivity ($\rho_{xx}$) and transverse resistivity ($\rho_{xy}$, i.e., the Hall resistivity) as shown by Eq. (3) and (4) below (190):

$$\rho_{xx} = \frac{(n_e\mu_e + n_h\mu_h) + (n_e\mu_e\mu_h^2 + n_h\mu_h\mu_e^2)B^2}{(n_e\mu_e + n_h\mu_h)^2 + \mu_e^2\mu_h^2(n_h-n_e)^2 B^2} \cdot \frac{1}{e} \qquad (3)$$

$$\rho_{xy} = \frac{(n_h\mu_h^2 - n_e\mu_e^2) + \mu_h^2\mu_e^2(n_h-n_e)B^2}{(n_e\mu_e + n_h\mu_h)^2 + \mu_h^2\mu_e^2(n_h-n_e)^2 B^2} \cdot \frac{B}{e} \qquad (4)$$

where $n_e$ ($n_h$) and $\mu_e$ ($\mu_h$) are the density and mobility of the electron (hole) band respectively. From the s*imultaneous* fitting for $\rho_{xx}(B)$ and $\rho_{xy}(B)$ using such a *two-band* model, both the densities and mobilities of the electron- and hole-bands can be obtained. Clearly, for a real system with more than one electron or hole band, this oversimplified model averages electron and hole bands and neglects any interband interactions. Although adding more bands to the above model is possible in principle, more accurate results may not be obtained with an overparameterized model. In fact, the two-band model already yields reasonable results for a variety of material systems, so it is reasonable to extend its application to topological semimetals.

Eq. (3) indicates that $\rho_{xx}$ tends to saturate at high fields where the $B^2$ terms dominate. Only when $n_e$ = $n_h$, i.e. the case of electron-hole compensation, $\rho_{xx} \propto B^2$ without saturation. Under such a circumstance, large MR is expected when mobility is high. Table 1 shows the mobilities of some representative topological semimetals acquired from two-band model analysis, which are indeed high, in the range of $10^3$ – $10^6$ cm$^2$ V$^{-1}$ s$^{-1}$. Such high *transport* mobility is consistent with the ultra-low residual resistivity at the zero-temperature limit (~n$\Omega$ to a few µ$\Omega$, see Table 1) as well as the high quantum mobility revealed by quantum oscillation studies (discussed in Section 3.2.2).

It is also worth noting that the two-band model, while widely used, just provides an approximate description for the magnetotransport properties of multiple-band materials. Firstly, Eqs. 3 and 4 are not applicable if there are "open orbits", which occurs when Fermi surface is not closed in the momentum space (190). Secondly, the negligence of inter-band interaction leads to an apparent contradiction – the carrier compensation appears to be necessary for the non-saturated MR according to Eq. 3, but the Hall resistivity expressed by Eq. 4 must be linearly dependent on the field when $n_e = n_h$, which is not true for most topological semimetals (e.g., see Fig. 2a). Thirdly, according to Eq. 3, even approximate electron-hole compensation should be able to lead to a quadratic or nearly-quadratic field dependence for $\rho_{xx}$. This indeed has been observed in a number of topological semimetals (48; 183; 191-193), but linear or even sub-linear MR has also been observed in a variety of samples (107; 171; 172; 174-180; 182; 183; 191; 194). The linear MR could be a classical effect due to strong current inhomogeneity (172) or have a quantum mechanical interpretation (195) (see Section 3.2.8), while the sub-linear MR may be attributed to the weak-antilocalization caused by strong SOC (196). With these considerations, it appears that the two-band model may be applicable only for a limited field range or at higher temperatures where quantum effects are not significant.

Although it might be challenging to obtain the precise value of carrier mobility for individual bands, the two-band model still provides an effective approach for the approximate description of magnetotransport properties of multi-band materials. This model successfully explains the extremely large MR arising from high mobility and approximate carrier compensation. Then a key question for topological semimetals is why Dirac/Weyl fermions have high mobility. This can be understood in terms of the energy band characteristics of topological semimetals. Given that the carrier mobility is determined by relaxation time $\tau$ and effective mass $m^*$ by $\mu = e\tau/m^*$, greater relaxation time and smaller effective mass favor higher mobility. As will be shown in Section 3.2.2, the cyclotron effective masses derived from quantum oscillations are indeed small for many topological semimetals, reaching as low as $0.02m_e$ ($m_e$: free electron mass) for some materials. Such massless behavior is naturally expected for ideal topological fermions since they are hosted by linearly dispersed bands crossing near the Fermi level, which requires zero mass in the Hamiltonian (11).

Greater relaxation time in topological materials may be associated with symmetry protection in many cases. For topological insulators, it has been well established that backscattering is forbidden by time reversal symmetry even though non-magnetic defects exist, thus resulting in longer relaxation time (197-201). In some topological semimetals, a strong suppression of backscattering due to non-trivial band topology has also been proposed (45), which leads to enhanced *transport* relaxation time. This is partially supported by the quantum oscillation studies which reveal a long *quantum* relaxation time in topological semimetals, as will be shown later in Section 3.2.2.

## 3.2 Landau quantization and quantum oscillations

In addition to the extremely large MR, another important phenomenon in the magnetotransport of topological semimetals is quantum oscillation (Figs. 2b and 2c), i.e. the Shubnikov–de Haas (SdH) effect. Quantum oscillations can also be probed in other measurements such as magnetization/magnetic torque [i.e. de Haas–van Alphen (dHvA) effect], thermoelectric power, ultrasonic absorption, etc. Quantum oscillations have been widely used for the study of the 3D topological insulators (202) and topological semimetals as it reveals key parameters for Dirac/Weyl fermions such as effective mass, quantum mobility, and more importantly, the Berry phase. In this section, we will present a review of quantum oscillation studies of topological semimetals.

### 3.2.1 Zeroth Landau level for relativistic fermions

Quantum oscillation theory for non-relativistic electrons has been well-established and documented in earlier textbooks and reviews (203; 204). Here we shall briefly recall the fundamental theory and put major emphasis on its extension to relativistic fermions. Quantum oscillation originates from the quantized cyclotron motion of charge carriers under magnetic fields, i.e., the Landau quantization of the energy states. With the conduction band splitting to Landau levels, the density of states at the Fermi level DOS($E_F$) becomes periodically modulated by magnetic field (more precisely, periodic in $1/B$), leading to periodic oscillations of physical quantities.

Figs. 3a and 3b show the textbook drawings of the Landau quantization for spinless (i.e., ignoring the Zeeman splitting) non-relativistic electrons with parabolic dispersion. The quantized Landau level (LL) energy is $\varepsilon_n = (n + 1/2)\hbar\omega_c$, where $\omega_c = eB/m$ is the cyclotron motion frequency, and the Landau level index $n = 0, 1, \ldots$ The energies of all LLs are field-dependent and evenly spaced by $\hbar\omega_c$, as shown in Fig. 3b. For the lowest LL, a finite zero-point energy $\hbar\omega_c/2$ exists, which is in analogy to the zero-point energy of a harmonic oscillator. In order to distinguish it from the exotic zeroth LL with field-independent zero energy for the relativistic fermions shown later, we rewrite the Landau level energy of non-relativistic electrons as $\varepsilon_n = (n - 1/2)\hbar\omega_c$ where $n$ becomes a non-zero integer (1, 2, …).

The Landau level quantization is completely different for the relativistic fermions with linear dispersion (Fig. 3c). Earlier studies on graphene have already established that (205; 206) the quantized energies of LLs for spinless *2D* Dirac fermions are:

$$\varepsilon_n = v_F \, \text{sgn}(n)\sqrt{2e\hbar|B||n|} \quad (n = 0, \pm 1, \pm 2 \ldots) \quad (5)$$

where sgn($n$) is the sign function and $v_F$ is the Fermi velocity. As illustrated in Fig. 3d, LLs are no longer equally spaced for relativistic fermions given $\varepsilon_n \propto \sqrt{|n|}$. Most strikingly, a field-independent zeroth ($n = 0$) LL locked at the band crossing point ($\varepsilon_0 = 0$) appears, which is a signature unique to 2D relativistic electron systems. Such a zero energy can be understood in terms of the Berry phase arising from the cyclotron motion of carriers in momentum space (206). The detailed theoretical background of Berry phase and its manifestation in transport measurements have been well-understood (202; 207-209). In short, Berry phase describes a geometrical phase factor of a quantum mechanical system acquired in the adiabatic evolution along a closed trajectory in the parameter space. Such a phase factor does not depend on the details of the temporal evolution and thus differs from the dynamical phase. A non-zero Berry phase $\phi_B$ originates from band touching point, such as Dirac nodes. Under magnetic fields, the cyclotron motion of Dirac fermions, i.e., closed trajectory in momentum space, induces Berry phase that changes the phase of quantum oscillations. Ideally, $\phi_B = \pi$ for an exact linear energy-momentum dispersion, and shifts from this value when the bands deviate from linear dispersion and/or Zeeman effect is strong (209; 210).

Before formulating the quantum oscillation for relativistic fermions by incorporating the Berry phase-induced phase shift, we should pay attention to the dimensionality of the investigated material systems. The Landau quantization of the 2D surface state of topological insulators is very different from that of the Dirac or Weyl fermions in 3D topological semimetals. Most topological semimetals reported so

far are 3D in nature (such as $Cd_3As_2$ (14; 127-130), $Na_3Bi$ (13; 126), TaAs-family (22; 23; 25; 27; 39; 86; 87; 211), etc) and 3D is necessarily required for a Weyl state (10). It is known that for non-relativistic electrons in 3D, the motion along the magnetic field direction is not quantized, leading to additional energy of $(\hbar k_z)^2/2m$ (where $k_z$ is the momentum along the magnetic field direction) for LLs:

$$\varepsilon_{n,k} = \frac{\hbar eB}{m^*}(n - \frac{1}{2}) + \frac{\hbar^2 k_z^2}{2m^*} \quad (n = 1, 2, 3, \ldots) \quad (6)$$

Similarly, an additional energy term due to un-quantized $k_z$ also occurs for 3D relativistic fermions:

$$\varepsilon_n = \upsilon_F \operatorname{sgn}(n)\sqrt{2e\hbar |B||n| + (\hbar k_z)^2} \quad (7)$$

Therefore, although the zeroth LL's energy is still field-independent, it is not strictly zero. It is also worth noting that Eq. 7 is valid for Dirac fermions with $n = 0, 1, 2, \ldots$ For Weyl fermions, the chirality is well defined due to the lifting of spin degeneracy, so Eq. 7 needs to be modified for the zeroth LL of Weyl fermions. As will be discussed in Section 3.4, the chiral zeroth LL leads to one important effect for Weyl fermions, i.e. the chiral anomaly.

### 3.2.2 The Lifshitz – Kosevich model for de Haas–van Alphen oscillations

For the perfect 2D case, the Landau bands are degenerate into sharp levels (Figs. 3b and 3d) and the motions of all electrons at the Fermi level are in-phase. For the 3D case, due to the additional energy related to un-quantized $k_z$ as shown in Eq. 6 and 7, different LLs overlap in energy space, leading to a mixture of Landau bands for particular energy (Figs. 3e and 3f) and a continuous energy spectrum. This is better illustrated in Fig. 3g: Landau quantization for 3D free electrons manifests as Landau cylinders along the magnetic field direction, so an equal energy surface (illustrated by dashed lines) intersects multiple Landau cylinders. This scenario is distinct from the 2D case (Fig. 3h). Therefore, different models have been derived for 3D and 2D quantum oscillations.

Here we start with the dHvA oscillation, because the magnetization is the derivative of the Gibbs thermodynamic potential $\Omega$ at constant temperature and chemical potential $\zeta$, $M = -\left(\frac{\partial \Omega}{\partial B}\right)_{T,\zeta}$, so that it directly reflects the Landau level spectrum. At the zero temperature limit, the oscillatory thermodynamic potential $\Omega$ due to Landau quantization for a 3D system can be expressed as (in CGS unit) (203):

$$\Omega_{osc} = \left(\frac{e}{2\pi c\hbar}\right)^{3/2} \frac{e\hbar B^{5/2}}{mc\pi^2 (\partial^2 S_{extr}/\partial k_z^2)^{1/2}} \sum_{r=1}^{\infty} \frac{1}{r^{5/2}} \cos[2\pi r(\frac{F}{B} - \gamma) + 2\pi\delta] \quad (8)$$

where $S_{extr}$ is the extremal Fermi surface cross-section area perpendicular to the magnetic field, $\partial^2 S_{extr}/\partial k_z^2$ is the Fermi surface curvature along the $k_z$ direction (i.e., the field direction) at the extremal cross section, and $r$ is the harmonic index. Considering several damping factors, the general formula of the magnetization oscillations for a 3D system, derived by Lifshitz and Kosevich (the LK formula) (203; 204; 212) is (in SI unit):

$$M_{osc}^{3D} = -\left(\frac{e}{2\pi\hbar}\right)^{3/2} \frac{S_{extr}}{\pi^2 m^*} \left(\frac{B}{|\partial^2 S_{extr}/\partial k_z^2|}\right)^{1/2} \sum_{r=1}^{\infty} \frac{1}{r^{3/2}} R_T R_D R_S \sin\left[2\pi r(\frac{F}{B} - \gamma + \frac{\delta}{r})\right] \quad (9)$$

$R_T$, $R_D$, and $R_S$ are the temperature-, field-, and spin-damping factors, which are respectively associated with the finite temperature corrections to Fermi-Dirac distribution function, finite relaxation time due to impurity scattering, and phase difference between the spin-up and spin-down subbands. These factors can be expressed as:

$$R_T = \frac{raT\mu/B}{\sinh(raT\mu/B)} \quad (10)$$

$$R_D = \exp(-\frac{raT_D\mu}{B}) \quad (11)$$

$$R_S = \cos\frac{r\pi g\mu}{2} \quad (12)$$

where $\mu$ is the ratio of effective cyclotron mass $m^*$ to free electron mass $m_0$. $T_D$ is the Dingle temperature that is relevant to the quantum relaxation time, and $a = (2\pi^2 k_B m_0)/(\hbar e) \approx 14.69$ T/K.

The *sine* term in Eq. 9 describes the oscillation with frequency $rF$ and phase factor $2\pi r(-\gamma + \frac{\delta}{r})$, where the fundamental frequency $F$ is linked to $S_{extr}$ by the Onsager relation $F = \hbar S_{extr}/2\pi e$. The determination of the phase factor is of particular interest for the quantum oscillation study of topological materials, since the Berry phase $\phi_B$ is connected to the phase factor via $\gamma = \frac{1}{2} - \frac{\phi_B}{2\pi}$. The Berry phase, which was not included in Lifshitz and Kosevich's original formalism (i.e., $\gamma = \frac{1}{2}$) (212), can effectively shift the phase of quantum oscillations (209; 210). The phase shift $\delta$ in Eq. 9, which is determined by the dimensionality of the Fermi surface, is 0 for the 2D case and ±1/8 for the 3D case. For the 3D case, $\delta = -1/8$ (+1/8) for maximal (minimal) cross-section for a 3D electron pocket (203; 204; 212) and vice versa for a 3D hole pocket.

Although most topological semimetals are 3D, there are also some materials with layered structure and thus display a quasi-2D electronic structure, such as ZrSiTe (156) and (Sr/Ba)Mn(Bi/Sb)$_2$ (143; 173; 177; 213). For a *perfectly* 2D system, the above LK formula has been modified by Shoenberg and others (203; 204; 214; 215):

$$M_{osc}^{2D} = -(\frac{e}{2\pi\hbar}) \frac{S}{\pi^2 m^*} \sum_{r=1}^{\infty} \frac{1}{r} R_T R_D R_S \sin\left[2\pi r(\frac{F}{B} - \gamma)\right] \quad (13)$$

with the same definitions for damping factors ($R_T$, $R_D$, and $R_S$) and phase factor $\gamma$ as the 3D model. The Fermi surface cross-section area become a constant for 2D, so $S_{extr}$ in 3D model (Eq. 9) is replaced by $S$, and the phase factor $\delta$ is zero. In addition to this phase difference, the oscillation amplitude (i.e. the prefactor of the summation in Eq. 13) and harmonic components ($r \neq 0$) are enhanced as compared to the 3D model.

It should be emphasized that the above 3D (Eq. 9) and 2D (Eq. 13) LK models are based on the assumption of constant chemical potential, which is appropriate for a 3D system because the electron energy spectrum is continuous as mentioned above. In this scenario, the lowest unoccupied state is always located at $E_F$ and independent of $B$ (i.e., chemical potential = $E_F$ for $T$ = 0 K). In contrast, the 2D Landau quantization gives rise to discrete energy levels, so the chemical potential, which is the minimum energy to add an electron into the system, is pinned to the highest occupied LL and hence also oscillates with ramping magnetic field. This chemical potential oscillation will affect the quantum oscillations. Furthermore, in real materials the interlayer coupling is not negligible in layered compounds, which is also not captured by Eq. 13. More comprehensive analyses can be found in Ref. (203; 204) and references therein.

In practice, the oscillation frequency(ies) $F$ can be directly resolved from the fast Fourier transform (FFT) of the oscillation pattern, and other important parameters including effective *cyclotron* mass, quantum relaxation time, and Berry phase can be obtained from the analyses with the LK formula. From FFT one can also clarify if the higher harmonic terms ($r >1$) with frequency $rF$ are significant or not. In principle, these terms attenuate quickly with $r^{-3/2}$ for a 3D system (Eq. 9) or $r^{-1}$ for a 2D system (Eq. 13), thus the quantum oscillations in real materials are usually dominated by fundamental frequencies ($r = 1$). If the oscillation contains only a single frequency without obvious harmonic frequency components, effective mass $m^*$ can be obtained from the fit of the temperature dependence of the oscillation amplitude $A_{osc}$ at a *fixed magnetic field* to the thermal damping factor $R_T$ in Eq. 10 (i.e. $M_{osc}(T) \propto R_T$). In normal metals with exact parabolic bands, the *band* effective mass is expected to be a constant despite the location of Fermi level. It can be easily shown that such band mass is equivalent to the *cyclotron* mass, which is defined as $m^* = \frac{\hbar^2}{2\pi}\left[\frac{\partial S}{\partial E}\right]_{E=E_F}$ within the semiclassical approximation, where $S$ is the extremal area enclosed by the cyclotron orbit in momentum space. Applying the same definition to the linearly dispersed bands with an isotropic Dirac cone, one can easily find that $m^*$ is connected to the Fermi vector $k_F$ and velocity $v_F$ with $m^* = \hbar k_F / v_F$. Thus $m^*$ should vanish when a Dirac point resides at $E_F$ (where $k_F = 0$), and increases when the Dirac point is shifted away from $E_F$. Such a trend has been observed in various Dirac materials (172; 216). Generally, $E_F$ is not too far away from the Dirac band crossing point in most of the known topological semimetals, so $m^*$ obtained from quantum oscillation is usually small, as summarized in Table 1.

With a known effective mass, the Dingle temperature which is associated with the quantum relaxation time can be extracted from the fit of the field dependence of the oscillation amplitude *at a fixed temperature* by the field damping factor $R_D$ in Eq. 11 (i.e., $M_{osc}(B) \propto R_D$). Because $T_D$ is included in the exponential term of $R_D$, the logarithm of the oscillation amplitude normalized by $B^{1/2}R_T$ (for 3D) or $R_T$ (for 2D) should have linear dependence on $1/B$ according to Eq. 11. Thus $T_D$ can be obtained from the slope of the linear fit of such a "Dingle plot". In practice, Dingle plots are non-linear in some cases where accurate $T_D$ cannot be obtained. This could be attributed to sample inhomogeneity, magnetic field inhomogeneity, "beating" oscillation pattern due to the existence of two very close frequencies, torque interaction at high fields if using torque magnetometry, etc (203).

From $T_D$ extracted from a Dingle plot, the quantum relaxation time $\tau_q$ can be derived via $\tau_q = \hbar/(2\pi k_B T_D)$. Because $\tau_q$ affects the oscillation amplitude exponentially (Eq. 11), strong dHvA oscillations present in low field ranges implies large $\tau_q$, which is generally the case for topological semimetals (Table 1). It is important to distinguish the quantum relaxation time from the transport relaxation time as mentioned in Section 3.1. While both arise from the scattering by static impurities and defects, these two quantities are essentially different (217; 218): The quantum relaxation time $\tau_q$ characterizes the quantum lifetime of the single-particle relaxation time of the momentum eigenstate, which determines the LL broadening of the momentum eigenstate by $\Gamma = \hbar/2\tau_q$, whereas the transport relaxation time $\tau_t$ is introduced in the classical Drude model and affects the Drude conductivity, $\sigma = ne\mu = ne^2\tau_t/m^*$. Given $\tau_t$ measures the motion of charged particles along the electric field gradient, it is largely unaffected by the forward scattering (i.e. small-angle scattering), which is distinct from $\tau_q$ that is susceptible to momentum scattering in all directions. Therefore, $\tau_t$ is usually larger or even much larger than $\tau_q$. Taking the form of the classical transport mobility $\mu_t = e\tau_t/m^*$, one can also define the quantum mobility by $\mu_q = e\tau_q/m^*$. Consequently, $\mu_q$ obtained from quantum oscillation is usually less than $\mu_t$ derived from magnetotransport, which has been observed in various topological semimetals as shown in Table 1.

In addition to nearly zero effective mass and high quantum mobility, non-trivial Berry phase is a key signature of relativistic fermions. As indicated above, it results in the zeroth LL, which is absent in the LL spectrum of non-relativistic electrons. In general, for a system exhibiting quantum oscillations with a single frequency, Berry phase $\phi_B$ can be determined from the LL index fan diagram, i.e., the plot of the LL indices $n$ vs. the inverse magnetic field $1/B$ (one example is shown in Figs. 4a and 4b). This method has been widely used in previous studies on topological insulators and a proper way to construct a LL fan diagram has been established though there had been some confusions in early studies (202; 219). We first consider a 2D situation. As shown in Figs. 3b, with ramping magnetic field, the LLs successively pass through $E_F$. Integer LL indices are assigned when $E_F$ lies at the middle of two adjacent LLs [i.e. minimum DOS($E_F$)], while half-integer indices are assigned when $E_F$ is right at the LL [maximum DOS($E_F$)]. For a LL fan diagram established with such a definition of LL index, the linear extrapolation of the linear fit of $n(1/B)$ to the $\frac{1}{B} \to 0$ limit must lead to $n = 0$ for non-relativistic electrons, but $n = 1/2$ for relativistic fermions due to the zeroth LL pined at the zero energy. This $n = 1/2$ intercept corresponds to an ideal Berry phase of $\pi$. For a 3D system, the phase of quantum oscillation is shifted by $2\pi\delta$ as mentioned above, so the linear extrapolation should intercept the $n$-axis at $\frac{\phi_B}{2\pi} - \delta$.

Therefore, proper assignment of LL indices is critically important to guarantee precise determination of Berry phase. Oscillations in *differential* magnetic susceptibility $\chi$ ($= \frac{dM}{dB}$) offers a straightforward approach to determine integer LL indices; that is, the minima of $\chi$ should be assigned with integer LL indices, since they correspond to minimal density of state at Fermi level, DOS($E_F$). This can be understood as follows: As indicated above, magnetization is equal to the derivative of the Gibbs thermodynamic potential $\Omega$ at constant temperature and chemical potential $\zeta$, $M = -\left(\frac{\partial \Omega}{\partial B}\right)_{T,\zeta}$. At zero temperature, $\Omega$ is indeed proportional to the total energy of electrons and modulated by magnetic field in the form of a cosine function (Eq. 8) (203). Given $\chi = \frac{\partial M}{\partial B} = -\frac{\partial^2 \Omega}{\partial B^2}$, $\chi$ and $\Omega$ would oscillate in phase when Landau quatization occurs with increasing magnetic field. Since the minima of $\Omega$ correspond to the minimal DOS($E_F$), minimal $\chi$ should be assigned with integer LL indices. Given $\chi = \frac{\partial M}{\partial B}$, if the oscillations of magnetization are used to establish a LL fan diagram, the minima of $M$ should be assigned with $n$-1/4 ($n$, integer number). With this approach, non-trivial Berry phase has been extracted from dHvA oscillations for several topological semimetals (156; 220-222).

It should also be emphasized that several factors can affect the value of Berry phase in topological semimetals. First of all, the Berry phase can deviate from an ideal value of π if the band dispersion is not perfectly linear (210). Second, the Zeeman effect, which has not been considered so far, also leads to a deviation of Berry phase obtained from a LL fan diagram (210). Therefore, the Berry phase determination using the LL fan diagram should be performed with caution for high field quantum oscillations or for materials with large *g*-factors such as $Cd_3As_2$ (172; 223) and ZrSiS (221). Furthermore, from the aspect of data analysis, reading Berry phase from a LL fan diagram may bear large uncertainty in some cases. Because Berry phase is determined by the intercept of the linear fit of $n(1/B)$, when low LL indices cannot be reached in experiments due to high oscillation frequency, a slight change in the slope of the linear fit can lead to a large shift of the intercept, thus resulting in a large uncertainty in the extracted Berry phase. Therefore, reaching low LL indices under high magnetic fields is necessary to obtain a reliable Berry phase from a LL fan diagram.

In addition to the magnetization measurements, dHvA oscillations can also be probed by torque magnetometry, since a magnetic moment $\vec{m}$ in a magnetic field is subject to a torque $\vec{\tau} = \vec{m} \times \vec{B}$. It is convenient to perform magnetic torque measurements on topological semimetals by using a cantilever (176; 224-230) to high magnetic field even up to 60 T. One drawback of the torque magnetometry is the "torque interaction", an instrumental effect due to the feedback of the oscillating magnetic moment on the cantilever position, which leads to artificial effects in quantum oscillations under high magnetic fields (203).

### 3.2.3 Shubnikov–de Haas oscillations

Besides dHvA oscillation, the resistivity oscillation, i.e., the SdH effect, is also widely used to study topological semimetals (46; 141; 171; 172; 174; 178; 179; 183; 191-193; 231-233). The extraction of Berry phase from SdH oscillations seems straightforward. Since the SdH effect also originates from Landau quantization, the non-trivial Berry phase associated with the zeroth LL also manifests itself by a phase shift in the SdH oscillation. As stated above, integer LL indices should be assigned when $E_F$ lies in the middle of two adjacent LLs and DOS($E_F$) reaches minima. The situation is less complicated in 2D integer quantum Hall systems (including the 2D surface states of the 3D topological insulators), in which the integer LL indices unambiguously correspond to the quantized Hall plateaus where the longitudinal conductance reaches minima ($S_{xx}=0$) due to the dissipationless edge state. The proper way to build a LL fan diagram from the SdH effect for topological insulators has already been discussed in a previous review article (202).

In the studies of topological semimetals, however, there have been controversies in constructing LL fan diagrams from the SdH effect. Various definitions for integer LL indices have been used in literature, including resistivity minimum (141; 178; 234; 235), resistivity maximum (171; 179; 183; 191; 193; 232; 233; 236-239), and conductivity minimum (143; 172; 213). At the first glance, it is natural to extend the above argument for the quantum Hall system to topological semimetals, except that the conductivity of topological semimetals cannot be directly measured through conventional transport experiments, but should be obtained through inverting the resistivity tensor, $\hat{\sigma} = \hat{\rho}^{-1}$. For in-plane (*x-y* plane) current $I$ and out-of-plane (z-direction) magnetic field $\boldsymbol{B}$ (i.e., a standard Hall effect setup with $\boldsymbol{B} \perp \boldsymbol{I}$) applied to a 2D system, the charge carriers undergo only in-plane motion and we have:

$$\hat{\sigma} = \begin{pmatrix} \sigma_{xx} & \sigma_{xy} \\ \sigma_{yx} & \sigma_{yy} \end{pmatrix} = \hat{\rho}^{-1} = \begin{pmatrix} \rho_{xx} & \rho_{xy} \\ \rho_{yx} & \rho_{yy} \end{pmatrix}^{-1} \qquad (14)$$

Here the resistivity tensor elements $\rho_{ij}$ ($i,j = x, y$) are defined as $\rho_{ij} = E_i / J_j$ ($E_i$, the electric field component along the $+i$ direction; $J_j$, the current density along the $+j$ direction), or equivalently, $V_i / I_j$ ($V_i$: voltage drop along $+i$ direction; $I_j$: current along $+j$ direction), which can be directly measured. In fact, from this definition, $\rho_{xx}$ and $\rho_{xy}$ are essentially the longitudinal and transverse (Hall) resistivity. Under the assumption of isotropic scattering rate for a given 2D material, it is easy to demonstrate $\rho_{xx} = \rho_{yy}$, and $\rho_{xy} = -\rho_{yx}$. Therefore, precise conductivity can be obtained from measured $\rho_{xx}$ and $\rho_{xy}$ via

$$\sigma_{xx} = \frac{\rho_{xx}}{\rho_{xx}^2 + \rho_{xy}^2}.$$

However, additional considerations must be taken for 3D topological semimetals. Although the integer quantum Hall effect also has a semiclassical interpretation based on Landau quantization, its underlying transport mechanism is distinct from the SdH effect due to its *nonlocal* character. As will be discussed in more details in Sections 3.2.7 and 3.5, the quantized Hall conductance plateaus and the zero longitudinal conductance are associated with the dissipationless edge channels. Such scale-invariant dissipationless edge conduction in quantum Hall systems is completely different from the transport in conventional diffusive systems where the resistance or conductance are associated with the sample dimensions and governed by the transport relaxation rate (i.e. the scattering rate). The scattering mechanisms in real materials can be very complicated. Fortunately, a semi-quantitative Lifshitz-Kosevich (LK) model that gives satisfactory descriptions for the SdH effect has been developed for 3D systems. The earlier transport theory has established that the scattering probability is proportional to the number of available states that electrons can be scattered into (47; 240), so it oscillates in concert with the oscillations of DOS($E_F$) and gives rise to SdH oscillations (203; 204). More explicitly, it has been shown that

$$\text{DOS}(E_F)_{osc} \propto (\frac{m^*B}{S_{extr}})^2 \frac{\partial M_{osc}}{\partial B}.$$ With this relation, the expression for conductivity/resistivity oscillation, i.e., the LK formula for the SdH effect, can be derived from the derivative of the magnetization oscillation (203; 204). Clearly, within the framework of this LK model based on the oscillation scattering rate, conductivity should exhibit *maxima* when the scattering rate reaches minima that occur at minimal DOS($E_F$). Given that integer LL indices should correspond to DOS($E_F$) minima as indicated above, the *maxima* of conductivity oscillation should be assigned with integer LL indices. However, it should be pointed out that this approach is based on the semi-quantitative model for the SdH effect (203). The scattering rate in a real material depends on a number of factors and can be very complicated particularly in multiband or anisotropic systems, which could lead the SdH oscillations to strongly deviate from the LK theory (204). As a result, a simple connection between the integer LL indices and the SdH oscillation extrema may be problematic in some cases. Therefore, to demonstrate non-trivial Berry phase, a better approach might be the oscillation of thermodynamic properties which are directly linked to Landau level energy spectrum, such as dHvA effect as discussed above.

In addition, the complication of the scattering rate in the SdH oscillation also leads to inconsistency between the SdH effect and dHvA effect. It has been shown that in some layered topological semimetals, dHvA oscillation is strong for arbitrary magnetic field directions, but SdH oscillation quickly attenuates when the magnetic field is tilted toward the current direction (221; 226; 232; 241; 242). In those materials, the stronger dHvA effect is also useful to distinguish the Zeeman splitting effect (221).

### 3.2.4 Multi-frequency quantum oscillations

The above discussions on LL fan diagram are applicable to quantum oscillations with a single frequency. However, multiple oscillation frequencies are often observed in most topological semimetals, such as the TaAs-family (179; 180; 182; 183; 191-193; 227; 243) and the *WHM* materials with PbFCl-type structure (*W* = Zr or Hf, *H* = Si, Ge, or Sn, *M* = S, Se, or Te) (156; 221; 222; 226; 228; 232; 233; 241; 242; 244; 245), etc. Given $F = \hbar S_{extr}/2\pi e$, the dependence of oscillation frequencies on the magnetic field orientation provides useful information of Fermi surface morphology. In the presence of multi-frequency oscillations, the method used to analyze effective mass, quantum mobility, and Berry phase differs from what is discussed for the single-frequency situation. The commonly used approach to obtain the effective masses is the fits of the FFT amplitudes for each frequency component to the thermal damping factor. Here the inverse magnetic field $\frac{1}{B}$ in $R_T$ (Eq. 10) is approximated by the average inverse field $\left\langle \frac{1}{B} \right\rangle$, defined as

$$\left\langle \frac{1}{B} \right\rangle = \frac{1}{2}\left(\frac{1}{B_1} + \frac{1}{B_2}\right),$$

where $\frac{1}{B_1}$ and $\frac{1}{B_2}$ are the upper and lower inverse fields used for FFT. However, although this method has been intensively used for a variety of topological semimetals [xxxx] and many other material systems [xxxx], one should be aware that the obtained effective mass may bear large errors because it is strongly depend on the range of the inverse magnetic field ($\frac{1}{B_1} \to \frac{1}{B_2}$) used for FFT. In Fig. xx we use the dHvA oscillation of ZrSiS as an example. The low field (up to 7T) dHvA oscillation of ZrSiS displays two frequencies (240T and 8.4T) with strong Zeeman splitting for the lower frequency band [xx] (Fig. xxa). As shown in Figs. xx b – c, the effective mass for both frequency bands increases strongly (by nearly a factor of 3) with narrowing the inverse field range for FFT. Such strong variation indicates the obtained effective mass may not be reliable. In ZrSiS, because the two oscillation frequencies are far apart (240T and 8.4T) [xx], it is easy to read the oscillation amplitudes for each frequency band directly from the oscillation pattern (Fig. xxa). Therefore, one can also extract the effective mass from the fits of the oscillation amplitudes rather than the FFT amplitudes, which is the standard method used for single frequency oscillations introduced in Section 3.2.2. This can provide an accuracy check for the effective mass obtained from the FFT amplitudes. As shown in Figs. xxd – e, the fits of the amplitudes at three different magnetic fields yield consistent effective masses for both bands (0.078 - 0.083 $m_0$ for the $F = 8.4$ band, 0.180 - 0.191 $m_0$ for the $F = 240$T band), in sharp contrast with the large variation seen for the fits for FFT amplitudes.

From the above example, errors in effective mass appear to be inevitable when extracted from the FFT amplitudes. For multi-frequency oscillations, if the frequencies are far apart or one has a dominant amplitude, it may be possible to obtain accurate effective mass by directly reading the oscillation amplitudes, as demonstrated in ZrSiS above. On the other hand, when frequencies are close to each other, several approaches can be used to reduce the error in effective mass [Antony Carrington, private communication]. First of all, as shown above for ZrSiS, the FFT effective mass trends to be close to the accurate value with narrow inverse field range. Second, it may be possible to use Fourier filter to reduce the problem to a single-frequency oscillation, in which the effective mass can be accurately obtained. Data near the two ends of the magnetic field range should be excluded after applying the Fourier filter because the ends effect could induce artificial signal. Third, the whole oscillation pattern can be fitted to the generalized multiband LK formula, with the assumption that the quantum oscillations of different bands are additive. To improve the accuracy one can fit all the multiple data sets at different temperatures and fields simultaneously. This is in principle always possible, but in practice the covariance between the parameters sometime could make it impractical. The combination of several different approaches, together with a simulation of the oscillation pattern using LK formula after obtaining these parameters, may be used to minimize the error in effective mass.

The Dingle temperature and Berry phase can be extracted through fitting the oscillation pattern to the generalized multiband LK formula. This method was previously used for the LaAlO$_3$/SrTiO$_3$ heterostructure (246), and was first employed for analyzing the SdH oscillations of TaP (Fig. 4c) in the study of topological semimetals (192) and then was proven to be effective in characterizing topological fermion properties for many other multiband topological semimetals (143; 156; 221; 226; 230; 245; 247-249). For the multiband LK fit, it is important to include all major frequency components, as well as the higher harmonic ($r > 1$ in Eqs. 9 and 13) terms if they are significant in the FFT spectrum, though there is a tradeoff for accuracy due to increased number of parameters. The spin damping factor $R_S$ is field-independent (see Eq. 12) and thus can be treated as a constant for the fit; it takes effects in modulating the amplitude for the harmonic component as it contains $r$. Furthermore, $R_S$ can be used to extract the Landé $g$-factor of a 2D/quasi-2D system via the "spin-zero" method; that is, the oscillation amplitude vanishes at some field orientation due to the interference of spin split Fermi surfaces. This provides an alternative method to evaluate the $g$-factor in addition to the direct measurement of the separation of the split oscillation peaks. Such analysis has been reported for ZrSiS (221) and WTe$_2$ (250).

### 3.2.5 Magnetic breakdown

Multiple oscillation frequencies usually result from multiple Fermi surface extremal cross-section areas perpendicular to the field. Additionally, charge carriers may tunnel from one cyclotron orbit to another, and jump back to the original one to form a bigger cyclotron orbit, hence leading to additional frequency(ies) equal to the *sum or difference* of two or more fundamental frequencies (203; 251). This phenomenon, called "magnetic breakdown", becomes more pronounced at high fields because the tunneling probability scales exponentially with the inverse field $1/B$ as $e^{-\alpha/B}$, where $\alpha$ is a material dependent parameter relevant to the $k$-space separation of the orbits (203). The additional frequencies ascribed to magnetic breakdown have been observed in high field quantum oscillation studies on several topological semimetals (171; 226; 252).

In type-II Weyl semimetals, the magnetic breakdown has been predicted to be associated with the "Klein paradox" proposed almost 90 years ago, which states that the tunneling barrier is nearly "transparent" for relativistic fermions when its height exceeds the electron's rest energy $mc^2$ (253). This relativistic effect is attributed to the positron or electron emission by a potential barrier when the barrier is sufficiently high (254-256). The matching between electron and positron wavefunctions across the barrier leads to high probability tunneling (257). However, the requirement of the high potential barrier ($\sim mc^2$) imposes a great challenge for the experimental observation of this phenomenon in particle physics. Fortunately, the (rest) massless relativistic fermions discovered in condensed matter provide a realistic platform given, in principle, there is no theoretical requirement on the potential barrier for massless relativistic fermions. The Klein tunneling has been demonstrated in graphene with a potential barrier created by a local gate (257; 258). A similar effect is also expected in topological semimetals with massless relativistic fermions. Recent theoretical work has predicted a momentum space counterpart of Klein tunneling in quantum oscillations for type-II Weyl semimetals (259). In the scenario of magnetic breakdown, quantum tunneling through different momentum space orbits naturally mimics real space tunneling of carriers (e.g. in graphene (257; 258)), which is expected to lead to unusual dependence of FFT amplitude on magnetic field orientation (259).

### 3.2.6 Quantum Oscillation due to Weyl orbits

The unusual surface Fermi arc is one distinct property of topological Weyl semimetals. For a Dirac semimetal whose Dirac node can be viewed as the superposition of two Weyl nodes with opposite chirality, its surface state exhibits two sets of Fermi arcs curving in opposite directions on two opposite surfaces, as shown in Fig. 4d. It has been predicted that under magnetic fields, electrons can transport on a cyclotron orbit which connects one surface Fermi arc to the opposite Fermi arc by coupling to bulk states (43; 260) (Fig. 4d). Such an unconventional "Weyl orbit" manifests itself by an additional frequency in quantum oscillations (Figs. 4e and 4f) with 2D character which can be verified by the measurement of the field orientation dependence of oscillation frequency (i.e., $F \propto 1/\cos\theta$). Quantum oscillations due to Weyl orbits exhibit anomalous properties such as a sample thickness-dependent oscillation phase shift. To observe such a Weyl orbit, it is necessary to reduce the sample size to suppress the contribution of the bulk states. This has been demonstrated in nanostructures of $Cd_3As_2$ (Figs. 3e-3f) (44; 261) and $WTe_2$ (262).

### 3.2.7 Other anomalous transport signatures originating from the zeroth LL

As has been indicated above, the field-independent zeroth LL of relativistic fermions leads to a phase shift in quantum oscillations from which the Berry phase can be inferred. In some layered topological semimetals, the zeroth LL has been probed more directly by several transport techniques such as quantum Hall effect (QHE) and interlayer tunneling.

The concept for QHE for 2D Dirac fermions has already been established for graphene and topological insulators (216; 263-265). Under a magnetic field, Landau quantization gives rise to quantized electron cyclotron orbits. Semiclassically, under sufficiently strong field, the electrons are pinned to these quantized small radii orbits which causes a bulk insulating state. However, electrons that are close enough to the edges cannot complete cyclotron motions but rather get bounced back by the edges. Considering the direction of Lorentz force, the reflected electrons have to move forward until being reflected by the edge again. This creates the so-called skipping orbit at the edge that carries current, i.e., the edge channel (Fig. 5a). Given the skipping orbit originates from cyclotron orbit, the number of the edge conduction channels is determined by the number of the quantized cyclotron motion states electrons can occupy, which is the number of the filled LLs below $E_F$. This gives rise to quantized Hall conductance of $G_{xy} = nG_0$ where $G_0 = e^2/h$ is the conductance quantum. In the language of band theory, the internal (bulk) of the 2D system is gapped when $E_F$ locates in between LLs. At the sample edge, the confining electrostatic potential that keeps electrons inside the sample bends the LLs upwards, as illustrated in Fig. 5a. The bent LLs which cross $E_F$ form the edge channels, giving rise to quantized Hall conductance. From the above edge channel-interpretation for QHE, it is clear that QHE is a *direct* manifestation of Landau quantization of electron energy states. This is in contrast with the SdH oscillation which arises from the oscillating scattering rate and thus is an *indirect* probe of LLs. In other words, QHE is a nonlocal transport phenomenon due to LLs, while SdH is a manifestation of LLs in local transport. Furthermore, QHE also has a topological interpretation which will be discussed in Section 3.5.

Given the existence of the field-independent zeroth LL pinned at the band crossing point (Figs. 3d and 3f), there is always an edge channel formed by the zeroth LL, as shown in Fig. 5a. Since the zeroth LL is evenly shared by both electrons and holes (Figs. 3f and 5a), the contribution of the zeroth LL to edge conduction is half of non-zero LLs', leading to the so called "half-integer quantization", i.e.

$$G_{xy} = G_0(n+\frac{1}{2}) \qquad (15)$$

This half-integer quantization can also be understood in terms of Berry phase of $\pi$ for relativistic fermions and has been observed in graphene (216; 263), zero-gap HgTe quantum wells (266), and 3D topological insulators (264; 265). In the real materials, an integer factor may be applied for $G_0$ due to the degeneracy, such as graphene with a factor of 4 orignating from spin and valley degeneracies (216; 263).

Given the difference in Landau quantization between 2D and 3D systems as mentioned in Section 3.2.1, it is challenging to probe *half*-integer QHE in 3D topological semimetals. One approach is to pursue their 2D nanostructures, but only integer QHE was observed so far in the nanostructures of $Cd_3As_2$ and $WTe_2$ (261; 267; 268), probably due to the quantum confinement effect which gaps the Dirac cone (267). We note that Masuda *et al* (177) reported a half-integer quantum Hall effect in a bulk Dirac semimetal $EuMnBi_2$ with layered structure (Fig. 5b). This material exhibits coexistence of two antiferromagnetic (AFM) orders, one formed by Mn sub-lattice and the other by the Eu sub-lattice. Application of a magnetic field induces a spin flop transition for the Eu AFM order, resulting in a canted AFM state, which reduces interlayer coupling significantly so that Dirac fermions generated by Bi square-net layers are more confined within the plane (i.e. quasi-2D) and exhibit signatures of half-integer QHE. As seen in Fig. 5c. $1/\rho_{xy}$ normalized by $1/\rho_{xy}^0$ ($\rho_{xy}^0$, the step size of successive plateaus) display quantized plateaus with half integers.

However, the quantum limit corresponding to $(1/\rho_{xy})/(1/\rho_{xy}^0) = 1/2$ could not be reached in this system, due to the fact that the canted AFM state of Eu sub-lattice can exist only in a limited field range.

In another structurally similar compound, YbMnBi$_2$, the zeroth LL was probed via interlayer transport (247). In this material, the Bi layers which host relativistic fermions are separated by the relatively insulating Yb-MnBi-Yb blocks, leading to a quasi 2D electronic state. As shown in Fig. 5d, given that two linear bands cross right at $E_F$ in this material (269), the 2D Landau quantization leads to the zeroth LL to be pinned to $E_F$ regardless of magnetic field strength. Therefore, increasing magnetic field leads to a monotonic increase in DOS($E_F$), which further enhances tunneling of electrons of neighboring Bi layers through the Yb-MnBi-Yb barrier when an interlayer electric field is applied. Because the 2D Landau quantization in YbMnBi$_2$ is governed by the magnetic field component perpendicular to the Bi plane, such exotic quantum tunneling of the zeroth LL carriers is sensitive to the magnetic field direction and can be detected in angular dependent magneto-transport such as *interlayer* magnetoresistance and *interlayer* Hall effect. For example, for the experimental setup shown in Fig. 5e, at low field when LLs are not well separated, LL broadening and thermal excitations smear out discrete LLs, which leads to conventional $(\sin\theta)^2$ dependence for the angular dependent *interlayer* resistance (AMR) (Fig. 5f, insert). In contrast, when the magnetic field is strong enough to establish the above quantum tunneling scenario, AMR reaches a broad minimum with $\theta$ being around 0° due to strong quantum tunneling, but sharply increases for the in-plane field orientation when 2D Landau quantization is suppressed. This causes a surprising strong peak centered $\theta = 90°$ in AMR, which can be well-fitted by the model that includes the tunneling of the zeroth LL's carriers (270) (Fig. 5f).

### 3.2.8 Beyond the quantum limit

When magnetic field is strong enough to push all LLs above $E_F$ except the lowest LL, all electrons are condensed to the lowest LL; such a state is generally referred to as a quantum limit. From this definition, one can find that the critical field needed to reach a quantum limit is at least comparable to the quantum oscillation frequency. The quantum limit is not accessible under a moderate magnetic field for most materials with high carrier density (i.e., large Fermi surface and large quantum oscillation frequency). A system under a quantum limit or an ultra-quantum limit may show unusual properties, which has been a long-standing topic of interest even for conventional materials. For instance, a fractional quantum Hall effect can occur near or in the ultra-quantum limit of a 2D electron gas (271). In topological semimetals, the dramatically enhanced degeneracy for the lowest LL, combined with the unique nature of relativistic fermions, may lead to some new exotic phenomena. Indeed, a mass enhancement in the quantum limit has been observed for $ZrTe_5$ (272). This was interpreted as the dynamic mass generation accompanied by density wave formation, which is due to the nesting of the zeroth LL driven by enhanced electron correlation (272). Another example of unusual transport in the quantum limit due to degeneracy enhancement is the aforementioned quantum tunneling of relativistic fermions in $YbMnBi_2$ (247). Because the zeroth LL is pinned at $E_F$ (269), the quantum limit can be reached in relatively low fields in this material (247).

Another phenomenon directly associated with the electron condensation to the zeroth LL in topological semimetals is the anomalous magnetization (224). The Landau quantization for a 3D Weyl semimetal yields energy spectra of:

$$\varepsilon_{n,k} = \begin{cases} \upsilon_F \, \mathrm{sgn}(n)\sqrt{2e\hbar |B||n|+\hbar^2 k_z^2}, & n \neq 0 \\ \chi \hbar \upsilon_F k_z, & n = 0 \end{cases} \qquad (16)$$

where $\chi = \pm 1$ represents the chirality of the Weyl points. At the quantum limit, the magnetization is purely contributed by the zeroth LL states, with $M_{n=0} = -\partial \varepsilon_{n=0,k} / \partial B$. Taking the derivatives of Eqs. 6 and 16, one can find that the magnetization per electron should saturate to a constant in a trivial metal, but vanish in the Weyl case. Therefore, one can expect a collapse of the magnetization for topological semimetals crossing the quantum limit. Indeed, the magnetic torque anomaly has been observed in NbAs, which can be quantitatively described by the topological character of the electronic dispersion (224).

High magnetic field may also lead to annihilation of a Weyl state. The recent studies on TaP have shown that the two counter-propagating chiral modes of the lowest LL (represented by $\chi = \pm 1$ in Eq. 16) may hybridize and open up an energy gap, leading to a magnetic-tunneling-induced Weyl node annihilation in TaP which manifests as a sharp reversal of the Hall signal (Fig. 6a) (273).

In addition to the above phenomena associated with the properties of the relativistic Dirac or Weyl fermions on the zeroth LL, new quantum states in the quantum limit regime have also been proposed (274; 275). For $ZrTe_5$ whose carrier density varies with different crystal growth techniques, its quantum limit can be reached under a very small magnetic field (~0.2 T) for low carrier density samples. In the quantum limit, surprising resistivity oscillations periodic in $\log(B)$ have been observed (Fig. 6b) (274), which are believed to be associated with the discrete scale invariance and formation of two-body quasi-bound state (274; 275).

Another old but intensively investigated transport behavior in the quantum limit is the linear MR. As discussed in section 3.1, the orbital MR stemming from the Lorentz effect should exhibit quadratic or nearly quadratic field dependence. In the quantum limit, however, it has been shown that MR grows linearly with $B$ (195). Such linear MR has been discovered in a number of materials (276-280) before the establishment of the theory for topological quantum states. In the recently reported topological semimetals, linear MR has been widely observed in many of them (45; 172; 173; 175; 234; 235; 281-283). However, linear MR for those materials begins to develop at a field much lower than the critical field needed to reach their quantum limits (45; 172; 173; 175; 234; 235; 281-283). Alternatively, it has also been proposed that the linear MR in $Cd_3As_2$ could also arise from spatial fluctuations of the magnitude and direction for local current density in disordered systems (172), which appears to be applicable for other topological semimetals with linear MR.

### 3.3 Intrinsic Anomalous Hall effect

In the last section we have intensively discussed the phenomena related to the Landau quantization and the zeroth LL in topological semimetals. As indicated above, the unique zeroth LL originates from the Berry phase of the band character of the relativistic fermions. In this section, we review another important phenomenon in magnetic topological semimetals, i.e. the intrinsic anomalous Hall effect (AHE), which also stems from Berry phase physics.

AHE, the enhanced Hall signal which couples with the magnetization of magnetic materials, has been intensively studied, as discussed in previous review articles (e.g. ref. (284)). Generally, the total Hall resistivity $\rho_{xy}$ in a ferromagnetic (FM) material has an anomalous contribution proportional to sample magnetization $M$ ($\rho_{xy}^{AH} = R_s M$) (284). The anomalous Hall resistivity can originate from the extrinsic mechanisms such as skew scattering (285) and side jumps (286) and/or the intrinsic mechanism due to the topological properties of bands (56; 287-289).

One important feature of magnetic Weyl semimetals is their *intrinsic* AHE. Such an intrinsic Hall component can be understood in terms of the Berry curvature $\vec{\Omega}$ of the electronic Bloch states, which leads to an anomalous electron group velocity perpendicular to the longitudinal electric field $[(e/\hbar)\vec{E} \times \vec{\Omega}]$(288). In a magnetic Weyl semimetal, a pair of Weyl nodes with opposite chirality can be seen as monopole sources of Berry curvature. In this case, the AHE is *purely intrinsic* and tunable by the separation of paired Weyl nodes (54). The intrinsic AHE current is dissipationless (55; 56; 284; 289), fully spin-polarized (289-291), and therefore of great potential for spintronic applications.

TRS-breaking Weyl state has also been predicted/established in many magnetic compounds. An incomplete list includes Co-based Heusler alloys Co$_2$XZ ($X$ = IVB or VB; $Z$ = IVA or IIIA) (95-99), half-metallic Co$_3$Sn$_2$S$_2$ (93; 94; 292), half Heusler compounds $R$PtBi (R = Gd and Nd) with AFM orders (108-110), and chiral antiferromagnets MnSn$_3$ and MnGe$_3$ (102; 103). The FM Co$_2$XZ compounds have been known as half metallic ferromagnets and some of them have Curie temperatures above room temperature, high spin polarization and large Seebeck coefficient (293; 294). It has been theoretically predicted that the locations of the Weyl points of these compounds in momentum space can be tuned by the magnetization direction (96; 97). These properties, together with the predicted giant anomalous Hall conductivity (98; 293), make these materials potentially useful for spintronic and thermoelectric applications. These predictions are still awaiting experimental verification. Recently, large intrinsic AHE and giant anomalous Hall angle have also been reported in FM $Co_3Sn_2S_2$ (94; 292), for which the existence of Weyl fermions has been demonstrated by the observation of the surface Fermi arcs (93).

The topological non-trivial states in half Heusler compounds have attracted significant attention even before the discoveries of topological semimetals (108; 295-297). The recent observations of chiral anomaly - a unique feature of Weyl fermions - together with band structure calculation suggest a magnetic field-driven Weyl state in AFM $R$PtBi (109; 110). Although different mechanisms including Zeeman splitting (109) and exchange field (110) have been proposed for the formation of a TRS-breaking Weyl state in these AFM zero gap semiconductors with quadratic band touching, intrinsic AHE associated with the magnetic field-driven Weyl state have been probed (Fig. 7a), with a very large anomalous Hall angle ~ 0.15 comparable to the largest observed in bulk ferromagnets (Fig. 7b) (110; 298).

The chiral antiferromagnets $Mn_3Sn$ and $Mn_3Ge$ have also been found to exhibit large anomalous Hall resistivity in the AFM ordered state, with a sharp and narrow hysteresis loop in magnetic field sweeps (Fig. 7c) (100; 101). Particularly, $Mn_3Sn$ is the first antiferromagnet that exhibits such a surprising large room temperature AHE (100). Further, remarkable anomalous behavior has also been observed in its Nernst effect (57). These anomalous transport features have been ascribed to a magnetic Weyl state, which was later demonstrated both theoretically (102) and experimentally (103).

In addition to the intrinsic AHE resulting from magnetic Weyl states, it is worth noting that the strong intrinsic AHE does not exclusively occur in magnetic Weyl systems. Other magnetic systems such as FM kagome metal $Fe_2Sn_3$ (299), FM spinel $CuCr_2Se_{4-x}Br_x$ (289), and magnetic semiconductors (288; 291) have also been reported to display intrinsic AHE.

### 3.4 Chiral anomaly

As a hallmark of Weyl semimetals, the Chiral anomaly is particularly important as it bridges the Weyl fermion physics in condensed matter and in high energy physics. Generally, the numbers of left- and right-handed Weyl fermions are conserved. This individual conservation of particles with opposite chirality is violated in the presence of parallel electric and magnetic fields. This effect, which was originally proposed in particle physics and called Adler-Bell-Jackiw effect or chiral anomaly (17), has been found to lead to exotic transport behaviors in condensed matter, i.e. negative longitudinal magnetoresistance, AMR narrowing, and planar Hall effect, which will be discussed in details below.

### 3.4.1   Chiral magnetic effect and negative longitudinal MR

Negative longitudinal MR (i.e. the increase of magnetic field parallel to the electrical current leading to a decrease of resistivity) related to chiral anomaly has been discovered in several topological semimetal systems as shown below. Chiral anomaly is indeed the manifestation of chiral magnetic effect - the generation of electric current under magnetic field induced by the chirality imbalance. The mechanism of this phenomenon has been well-established (10; 11; 52; 53). Here we first give a brief overview on its relevant physics. We first consider the quantum limit where only the zeroth LL is occupied. As described in Eq. 16 and illustrated in Fig. 8a, the 3D Landau quantization of a Weyl semimetal leads to counter-propagating zeroth LLs for a pair of Weyl cones, which disperse only along the magnetic field direction. This direction is also the direction for electrons to have coherent motion when an external electric field $\mathbf{E}$ is applied. Such electric field-driven motion leads to electron pumping between Weyl nodes with a rate $\propto -\mathbf{E}\cdot\mathbf{B}$ (10; 11; 53), which results in imbalanced population of carriers between the two zeroth LLs of the paired Weyl cones. As a result, the chirality becomes imbalanced. In condensed matter, this charge pumping process is finally relaxed by inter-Weyl node scattering and a steady state is reached, with a "chiral current" $j_c \propto B\mathbf{E}\cdot\mathbf{B}\tau_{int}$ where $\tau_{int}$ is the inter-node relaxation time (10; 11; 53). Clearly, this chiral current contributes to negative MR when $\mathbf{E}//\mathbf{B}$. Aside from this quantum mechanical interpretation based on only the zeroth LL, a semiclassical approach based on Boltzmann equation also yields the same result; with this approach, it can also be generalized to the semiclassical regime that involves multiple Landau levels (10; 11; 53).

Although the negative longitudinal MR originating from chiral magnetic effect occurs in both the quantum limit and semiclassical regime, the actual field dependence of MR can be material dependent. Generally, the negative MR is expected to be linearly dependent on **B** in the quantum limit, while being $\propto B^2$ in the low field range. But the real situation can be more complex if the inter-node scattering which relaxes the chiral charge pumping becomes field dependent. This is possible in the quantum limit at high field as will be shown below. In real materials, the situation can be further complicated by the positive orbital MR due to Lorentz effect, which is determined by the magnetic field component perpendicular to current as discussed in Section 3.1. Ideally, such positive orbital MR should vanish when **E**//**B**, but it has been shown that finite orbital MR may arise from anisotropic Fermi surface for **E**//**B** (300). Considering such orbital effects, the longitudinal MR may show quadratic field dependence in the low field range, but becomes negative when the chiral magnetic effect dominates.

It is also worth noting that the chiral magnetic effect is not limited to the situation of exact **E**//**B**, since the chiral charge pumping rate is finite for non-orthogonal electric and magnetic fields. Therefore, negative MR may be observed in a range of field orientation angle and vanishes when it is compensated by the positive orbital MR component which is determined by the transverse magnetic field component. In fact, if the negative MR is too sensitive to field orientation (e.g., it disappears when the magnetic field is deviated by 1 or 2 degree from the parallel direction), it may suggest a classical origin of "current jetting", which will be discussed below.

Chiral magnetic effect was first observed in the Dirac systems such as $Bi_{0.97}Sb_{0.03}$ (301), $Na_3Bi$ (105), $Cd_3As_2$ (Fig. 8c) (45; 106), and $ZrTe_5$ (Fig. 8d) (107) before the experimental discovery of Weyl semimetals. This can be attributed to the fact that the Dirac point in a 3D Dirac semimetal can be viewed as superposition of two paired Weyl nodes with opposite chirality. Such two overlapping Weyl nodes can be separated in momentum space by magnetic field which breaks time-reversal symmetry (Fig. 8b). Half Heusler $R$PtBi is another group of materials which exhibits magnetic field-induced chiral magnetic effect (109; 110). As mentioned in Section 3.3, these materials are zero gap semiconductors and their Weyl points are believed to be caused by the external field-induced Zeeman splitting (109) or exchange field from 4f electrons (110). It has been proposed that their Weyl points can be induced for any magnetic field orientation, and the induced Weyl points do not necessarily reside on the axis parallel to the field (104). For these field-induced Weyl states, the separation of Weyl points in momentum space may be dependent on magnetic field, so the negative longitudinal MR could display non-universal field dependence. For example, a quadratic field dependence of negative MR anticipated for a non-quantum limit regime has been observed for most of the above materials (107; 110; 301), however a saturation behavior is seen in Na$_3$Bi (Fig. 8c), which is attributed to the field-dependent inter-node relaxation time in the quantum limit (105).

Since the experimental discoveries of Weyl semimetal state in materials such as TaAs-class (type I) (22; 23; 25; 27; 39; 86-92) and (W/Mo)Te$_2$ (type II) (28; 111-122), many groups have reported observation of negative longitudinal MR in those materials and attributed it to chiral magnetic effect (109; 110; 179-181; 183; 192; 225; 302; 303). Although chiral anomaly is usually viewed as a smoking gun evidence for a Weyl state, one must be cautious before attributing the observed negative longitudinal MR to chiral anomaly, since a classical effect, current jetting, can also lead to negative longitudinal MR (47). Current jetting is simply due to the rule that the current flows predominately along high conductance direction. Once there exists large conductance anisotropy, equipotential lines are strongly distorted and thus the current forms "jets". For materials with large transverse MR, which is the case for most Dirac and Weyl semimetals, magnetic field causes very strong conductance anisotropy between the along-current and perpendicular-to-current directions. Therefore, with increasing magnetic field, the voltage drop between voltage contacts may even decrease for asymmetric point-like electrical contacts and irregular sample shape, leading to negative longitudinal MR (10; 304; 305). To minimize such a classical effect, it is important to use a perfect bar shape sample with a large aspect ratio and well separated, symmetric voltage contacts.

Current jetting is also expected to be weak in materials with small transverse MR (e.g. GdPtBi (304)) due to reduced conductance anisotropy under magnetic fields. More comprehensive discussions on current jetting effect in topological semimetals can be found in Refs. (304; 305).

For type-II Weyl semimetals such as (W/Mo)Te$_2$ (28; 111-122), chiral anomaly shows a different situation. Given their strongly titled Weyl cones, Landau quantization sensitively depends on the orientation of magnetic field and the Landau spectrum is gapped for some field directions. Therefore, their negative longitudinal MR is strongly anisotropic (28; 306; 307); this has been observed in WTe$_2$ (302; 303). Further studies also found that in the classical limit characterized by $\omega_c \tau \ll 1$ (as opposed to the quantum limit or semiclassical limit where $\omega_c \tau \gg 1$, where $\omega_c$ is the cyclotron frequency and $\tau$ is the transport relaxation time), negative longitudinal MR in Type-II Weyl semimetal becomes isotropic, similar to that in Type-I semimetals (303; 308).

### 3.4.2 Planar Hall effect (PHE)

In addition to generating negative MR in longitudinal transport, the chiral anomaly also leads to a non-trivial transverse (Hall) signal under in-plane magnetic field (Fig. 9a). Intuitively, an in-plane Hall signal is not expected under in-plane magnetic field due to the absence of electron accumulation on the sample edges. However, it has been shown that in-plane Hall voltage can be generated in presence of coplanar electric and magnetic field (Fig. 9a) due to chiral anomaly, leading to the so-called "planar" Hall effect (PHE) (309-315).

PHE is in fact a well-known phenomenon observed in ferromagnets which is due to the resistivity anisotropy caused by anisotropic magnetization (316). Although topological semimetals share the same in-plane angular dependence in Hall resistivity $\rho_{xy}$ with ferromagnets, PHE in topological semimetals occurs in the absence of magnetic order, with a significantly enhanced amplitude (309; 310). With coplanar electric and magnetic fields, the transverse resistance $\rho_{xy}$ of PHE is found to be (309):

$$\rho_{xy} = \frac{(\rho_\parallel - \rho_\perp)}{2}\sin 2\varphi \qquad (17)$$

where $\rho_\parallel$ and $\rho_\perp$ correspond to the resistivity with current flowing along and perpendicular to the direction of the magnetic field respectively, $\varphi$ is the angle between the current flow and magnetic field orientation (Fig. 9a). As mentioned in Section 3.2, in the Drude model the orbital MR for $B//I$ is strictly zero unless a multiband effect is involved. Therefore, the $\rho_\parallel - \rho_\perp$ represent the resistivity anisotropy caused by chiral anomaly.

In the experimental studies on Dirac and Weyl semimetals, an abnormal Hall signal under in-plane magnetic field was first reported in ZrTe$_5$ (317). A strict $\sin 2\varphi$ dependence was later observed in a number of materials including ZrTe$_5$, Cd3As$_2$, GdPtBi, WTe$_2$, and VAl$_3$ (311-315). With rotating in-plane (Fig. 9a) and out-of-plane field (Fig. 9b), respectively, the two-fold anisotropy of the PHE (Fig. 9c) clearly differs from the one-fold symmetry seen for the conventional Hall effect (Fig. 9d) (313). Unlike the conventional Hall effect, PHE does not satisfy antisymmetry, i.e., $\rho_{xy} \neq -\rho_{yx}$. This is because that PHE does not originate from the Lorentz force (309; 310).

### 3.4.3 Angular dependent *interlayer* resistance (AMR) narrowing

With the above definition of $\rho_\parallel$ and $\rho_\perp$, the longitudinal resistivity can be expressed as (309):

$$\rho_{xx} = \rho_\perp + (\rho_\parallel - \rho_\perp)\cos^2\varphi \qquad (18).$$

Another unusual property that can be derived from Eq. 18 is the narrowing of the AMR peak at high magnetic field (309). For simplicity, the magnetoconductivity with sweeping in-plane angle $\varphi$ may be expressed as $\frac{1}{\rho_{xx}(B,\varphi)} - \frac{1}{\rho_{xx}(0,\varphi)}$ (a stricter process requires tensor conversion). At a small angle, the angular dependence of magnetoconductivity has a Lorentzian profile with angular width (309):

$$\Delta\varphi \approx (\frac{\varepsilon_F}{\hbar v_F / l_B})^2 \sqrt{\frac{\tau}{\tau_c}} \quad (19),$$

where $l_B = \sqrt{\hbar/eB}$ is the magnetic length, $\tau_c$ is the relaxation time for chiral charge diffusion, and $\tau$ is the conventional momentum relaxation time. At low fields, Landau levels are wiped out by energy level broadening and thermal excitation. In this case, parameters involved in Eq. 19 are field independent except for $l_B$, indicating a narrowing of angular width with $B$ that has been observed in Na$_3$Bi (Figs. 8e-8f) (176). When a strong magnetic field drives the system to the quantum limit, the field dependence of each parameter in Eq. 19 leads to the saturation of $\Delta\varphi$, as shown in Figs. 9e-9f (176).

### 3.5 Quantum Hall states in the 2D limit

In the 2D limit, one intriguing aspect of topological semimetals is the potential to generate various quantum Hall states. In Section 3.2.8 we have mentioned the QHE in the 3D layered topological semimetal EuMnBi$_2$, which is caused by the formation of 2D electronic states due to restriction of electron motion in the 2D Bi-plane (177). Here we discuss another two quantum Hall states in the 2D limit which are of potential applications in electronics and spintronics: the quantum spin Hall insulator (QSHI, i.e., 2D topological insulator) and the quantum anomalous Hall insulator (QAHI).

The 2D quantum Hall states for both non-relativistic and relativistic electrons reflect the fundamental topological properties of materials. For example, the integer quantum Hall effect, an old phenomenon that was well understood in terms of the Landau quantization, now has a topological interpretation based on the topological invariant of Chern number, which opens the field of topological electronic states in condensed matter. As shown in Fig. 10a and mentioned in Section 3.2.7, an integer quantum Hall system under sufficiently strong fields is characterized by an insulating bulk state with electrons pinned to quantized small radii orbits, and the conducting, dissipationless chiral edge state formed by skipping orbits. The superposition of two copies of time-reversal integer quantum Hall systems in the quantum limit leads to QSHI, i.e., the 2D topological insulator, which displays a pair of counter-propagating, spin-polarized edge states due to spin-orbit locking (Fig. 10c). Apparently, the magnetic field which is necessary to produce an integer quantum Hall system is no longer needed for QSHI (76; 79), as it is cancelled out when bringing the time-reversal copies of integer quantum Hall systems together. Another modification of the integer quantum Hall system that does not require an external magnetic field is the QAHI, in which the spontaneous magnetization leads to the dissipationless chiral edge state (Fig. 10b) and the formation of Landau levels is not required (76; 79).

QSHI and QAHI also provide significant insights into topological physics beyond the simple modification of the integer quantum Hall system (76). The QAHI as well as the integer quantum Hall system are essentially the 2D Chern insulators characterized by non-zero Chern number, in contrast with the trivial insulator with $C = 0$. With TRS, the Chern number must vanish, but another topological invariant, the $Z_2$ number, can be introduced to clarify the 2D insulators, becoming 0 for trivial insulator and 1 for the symmetry-protected topological insulator (QSHI) (318). Simple stacking of these 2D building blocks leads to 3D "weak" Chern insulator or "weak" topological insulator which is not robust against disorder (319). It is also possible to extend the topological classification of QSHI to 3D and create the "strong" 3D topological insulator (319). However, the extension of the 2D Chern insulator to 3D cannot produce a "strong" 3D Chern insulator. Instead, this results in a metallic phase – the topological semimetal (76). The above discussions show how the quantum Hall system, QSHI, QAHI, 3D topological insulator, and topological semimetals are closely connected in terms of the topological properties, which implies the possibility of conversion between these states.

From the experimental aspect, QSHI and QAHI are expected to display unusual nonlocal transport (320; 321). The resistance or conductance of conventional diffusive systems are dependent on the dimensions of the sample and determined by the local resistivity or conductivity (Ohm's law). However, in quantum Hall systems, due to the scale-invariant dissipationless edge conduction, the transport is non-local and the concepts of resistivity or conductivity are thus meaningless in this case. The Hall conductance can be obtained from Chern number $C$ by $G_{xy} = Ce^2/h$; a half-quantized Hall conductance is also expected for massless relativistic fermions as mentioned in Section 3.2.8 (Eq. 15). For a QSHI, $G_{xy} = 0$ due to $C = 0$ in a TRS system, which can be attributed to the fact that the pair of time-reversed chiral edge states cancels each other (Fig. 10c). For the longitudinal conductance $G_{xx}$, the measurement results strongly depend on the configuration of the contact electrodes. This is because an ideal contact attached to the edge of the sample acts as a reservoir which draws and emits electrons from and to the edge channels. The spin information of an electron is smeared out during this process. For an integer quantum Hall system or a QAHI, the edge state is chiral (Figs. 10a and 10b) and the electrons emitted from the contact have to flow along the same direction, which should lead to zero longitudinal conductance and hence zero longitudinal resistance according to resistivity and conductivity tensor conversion. However, for a QSHI with time-reversed spin-polarized edge states, the spin of the emitted electrons has half probability to be reversed, corresponding to the back-moving edge channel with opposite spin. Therefore, a finite resistance depending on the number and configuration of contacts can be expected (320; 321).

### 3.5.1  Material realizations for QSHI and QAHI

QSHI has been proposed in the monolayer form of the layered 1T'- transition metal dichalcogenides $MX_2$ ($M$ = W, Mo; $X$ = S, Se, Te) (72) and $WHM$ (322). The structure of monolayer $MX_2$ is formed from the stacking of X-M-X layers, with its physical properties being determined by the type of stacking. The hexagonal "H" structure with ABA stacking (Fig. 10d) results in the well-known direct band gap semiconductors (323). For the rhombohedral "1T" phase with ABC stacking (Fig. 10e), the structure is found to be unstable and undergoes a spontaneous lattice distortion to the 1T' phase (Fig. 10f), which consequently leads to a QSHI state in presence of SOC (72). QSHI state in monolayer 1T'-$MX_2$ was first demonstrated in $WTe_2$ as this material naturally possesses the 1T' structure in the bulk form. Transport (74; 75) and spectroscopic (73) evidence of the QSHI state in $WTe_2$ monolayers prepared using mechanical exfoliation or molecular beam epitaxy (MBE) growth have been reported. For example, with sweeping the gate voltage, a conductance plateau associated with the 1D edge state of QSHI is observed in a $WTe_2$ monolayer (Fig. 10g), but absent in bi-layer or few-layer samples (74; 75). More importantly, the temperature at which the conductance plateau starts to develop is as high as 100 K (Fig. 10g), which is greatly higher than the operating temperature of other well-established QSHIs in semiconductor quantum wells (324) and could be ascribed to the large bulk band gap of the 1T'-$WTe_2$ monolayer (which was predicted to be 100 meV (72) and found to be 55±20 meV for MBE-grown samples (73)). This implies a great potential for practical device applications. Furthermore, it has been proposed that the horizontal electric field breaks the inversion symmetry and induces strong Rashba splitting of the bands near the $E_F$, which closes the bulk gap at some critical electric fields. Such gap closing leads to a topological phase transition to a trivial phase, which occurs very rapidly and thus can be used for topological field effect transistors (72).

The tetragonal layered *WHM* compounds have also been predicted to become a QSHI in the monolayer form (322). Different from $WTe_2$ which is a type-II topological Weyl semimetal in the bulk form (28; 111-113; 117), the bulk *WHM* is predicted to be a weak topological insulator formed from the stacking of QSHIs (322; 325), which is a long-sought topological quantum state (326). In *WHM*, the $C_{2v}$ symmetry ensures nodal-line crossings near the $E_F$ in the absence of SOC, but this symmetry cannot prevent SOC gap opening (154). Because that the Fermi level crosses the gapped cones and the band dispersion is extremely linear over a wide energy range, *WHMs* have been established as topological nodal-line semimetals (78; 85; 154). To realize the predicted QSHI state, one possible route is to exfoliate the bulk

*WHM*s to their monolayers. Although the interlayer coupling in *WHM* is not van der Waals-type (322; 327), the weak coupling strength in some *WHMs* allows for the mechanical exfoliation, as has already been demonstrated (156). One possible advantage of using *WHM*s as a platform for searching QSHI is the variable SOC gap with various combinations of *W*, *H*, and *M* (226), which offers the opportunity to design different QSHIs.

As mentioned above, a QAHI is in principle similar to the integer quantum Hall system, but it occurs without an external magnetic field and Landau levels (76; 79), which carries great promise for possible applications in spintronics. Furthermore, QAHI also provides a promising platform for the creation, manipulation, and utilization of Majorana fermions, the hypothetical particles that are their own antiparticles (328; 329). The QAHI state was first experimentally demonstrated in magnetically doped topological insulators (330-332). However, it has so far been realized only at very low temperatures (< 1 K) (330-332). Room temperature QAHE, if realized, will have the potential to revolutionize information technology through dissipationless spin polarized chiral edge transport in spintronic devices. Recent studies have revealed a new possible route to the realization of high-temperature QAHI: 3D FM Weyl semimetals can evolve into large gap QAHI when the dimensionality is reduced from three to two dimensions, due to the confinement-induced quantization of low energy states (21). One possible candidate material is $HgCr_2Se_4$ (21), which is awaiting experimental verification. In addition to these two approaches, there are also other proposals on the realization of QAHI, which can be found in Ref. (76).

## 4. Summary and Perspective

In summary, we have reviewed distinct electronic transport phenomena associated with the non-trivial band topology in different types of topological semimetals and discussed how to extract the fundamental properties of Dirac/Weyl fermions such as effective mass, quantum mobility and Berry phase from dHvA or SdH quantum oscillation measurements. From our discussions given in this review, it can be seen that topological semimetals exhibit a rich variety of exotic properties which are not seen in non-relativistic electron systems. This includes chiral anomaly and planar Hall effect in WSMs, intrinsic anomalous Hall effect in time reversal symmetry-breaking WSMs, quantum oscillations due to Weyl orbits and AMR peak narrowing under high magnetic fields in DSMs, half-integer quantum Hall effect and quantum tunneling of the zeroth LLs in layered magnetic DSMs, vanishing magnetization and dynamic mass generation in the quantum limit of DSMs/WSMs. We have made efforts to discuss how these properties are connected with the non-trivial band topology though the mechanisms for some of them have not been fully understood. Furthermore, we have also discussed how DSMs/WSMs are linked with QSHI and QAHI and how these two quantum Hall states can be approached by reducing NLSMs/FM WSMs to 2D thin layers.

As has been summarized in previous reviews (10; 11), one challenge in this field is the experimental realization of ideal model systems like "graphene" (10) or "Hydrogen atom" (11) for various types of topological semimetal phases. An ideal model system should contain only the topological band(s), with the same type of Dirac or Weyl points being symmetrically related, located at the Fermi energy, and well-separated in momentum space. For the material aspect, it should be stable in ambient environment and have minimal defects (10; 11). As noted above, topological semimetals discovered so far are probably just the tip of an iceberg. Given that topological semimetals can be predicted by band structure calculations, we believe many new topological semimetal phases and candidate materials will be discovered and some of them may serve as model systems. There have been recent breakthroughs in topological phase screening and database development for topological quantum materials (37; 325; 333-336). With new simple model systems, exotic phenomena arising from the non-trivial bands will not be masked or interfered by the contributions from the trivial bands, and novel knowledge in various topological semimetal phases can be further revealed.

Topological quantum materials have stimulated great interest, not only because of their connection with high energy particle physics, but also due to their great potential in future technology applications. As discussed above, both QSHI and the QAHI can be obtained by reducing the dimension of DNLS/FM WSMs to 2D and these two states can support dissipationless transport through their topological spin polarized edge states. Therefore, they carry great promise for applications for spintronic devices and quantum computation. Although both QSHI and the QAHI have been demonstrated experimentally, these states occur only in the low-temperature range at present. Pushing their operation temperature to room temperature is another great challenge in the field. Again, this requires discoveries of new topological materials with better properties. Realization of this goal requires integrated efforts of theoretical modeling, computation, synthesis, characterization and device demonstrations.

**Acknowledgement**: JH is supported by the U.S. Department of Energy (DOE), Office of Science, Office of Basic Energy Sciences under Award DE-SC0019467. ZQM is supported by the US National Science Foundation under grant DMR1707502. NN is supported by the U.S. Department of Energy (DOE), Office of Science, Office of Basic Energy Sciences under Award Number DE-SC0011978. We thank Prof. Antony Carrington from Bristol University for the informative discussions on the effective mass for multi-frequency oscillations.

# Figures

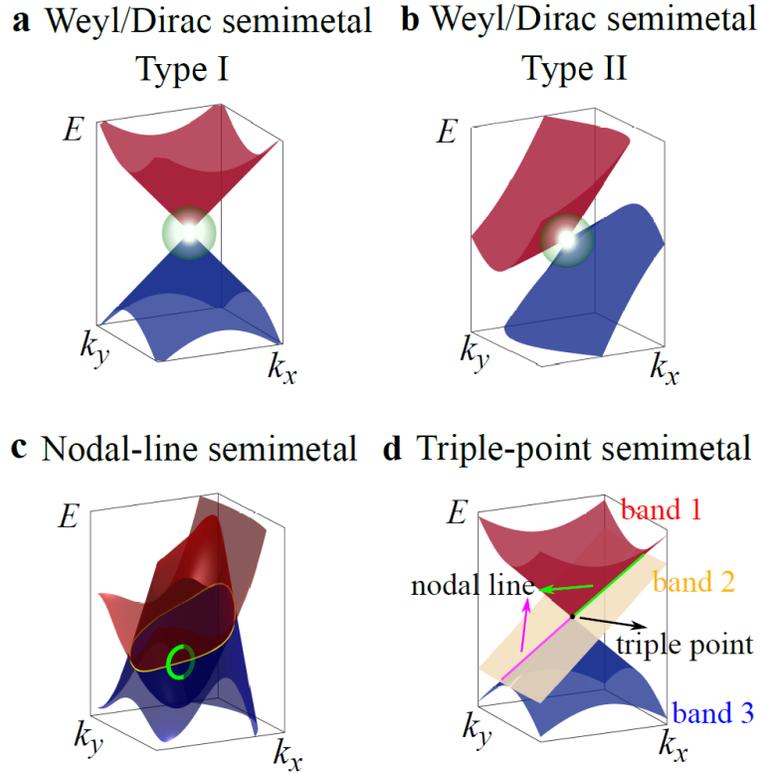

**Figure 1: Schematic band structure of different types of topological semimetals.** (a) Type-I Weyl/Dirac semimetal. The degeneracy of a Weyl point is half of that of a Dirac point. On a two-dimensional (2D) closed surface (the green surface) that encloses the Weyl node in $k$ space, the band structure is fully gapped and therefore allows for a topological invariant to be defined (19). Specifically, the topological invariant for a Weyl node is a chiral charge, which corresponds to the Chern number associated with the 2D closed surface. (b) Type-II Weyl/Dirac semimetal. At the energy of the type-II Weyl/Dirac node, the constant energy contour consists of an electron pocket and a hole pocket touching at the node. (c) Nodal-line semimetal. The conduction and valence bands are degenerate on a 1D closed loop shown as the green circle in the Brillouin zone. The topological invariant of the nodal line is a winding number $w$, which is defined as the line integral of the Berry connection along a closed loop shown as the green circle that interlinks the nodal-line. (d) Triple-point semimetal. Three singly degenerate bands cross at discrete points,

the triple-points. The triple-point can also be viewed as the meeting point between two nodal lines along the $k_y$ axis.

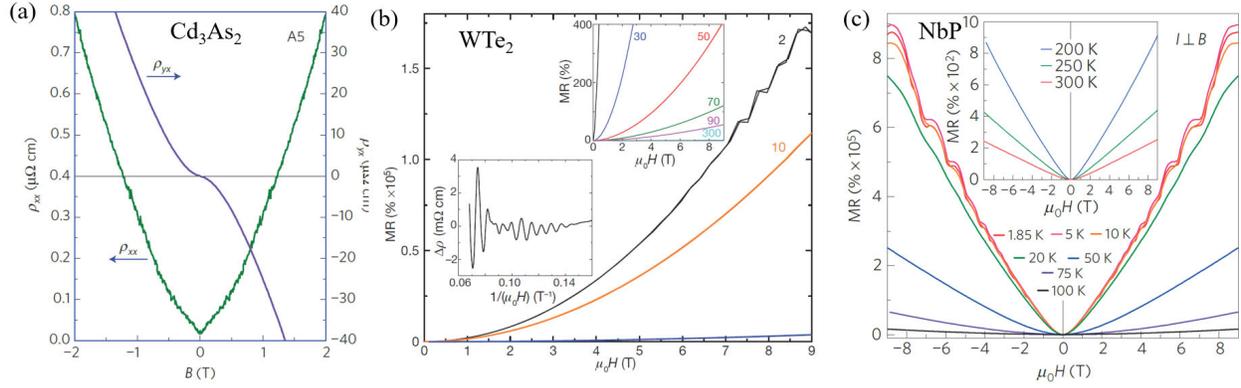

**Figure 2: Magnetoresistance.** (a) Magnetic field dependence of the longitudinal ($\rho_{xx}$) and transverse (Hall) resistivity ($\rho_{xy}$) for Cd$_3$As$_2$ (45). (b) MR normalized by the zero-field resistivity for WTe$_2$ at the temperatures of 2 K and 10 K. SdH oscillation is seen for the $T$ = 2 K data. Upper inset: MR at higher temperatures; Lower inset: oscillatory component of the resistivity oscillation, obtained by subtracting the smooth MR background. (48) (c) MR normalized by the zero-field resistivity for NbP at various temperatures. SdH oscillation is seen at temperatures ($T$ < 10K). Inset: MR at higher temperatures. (46)

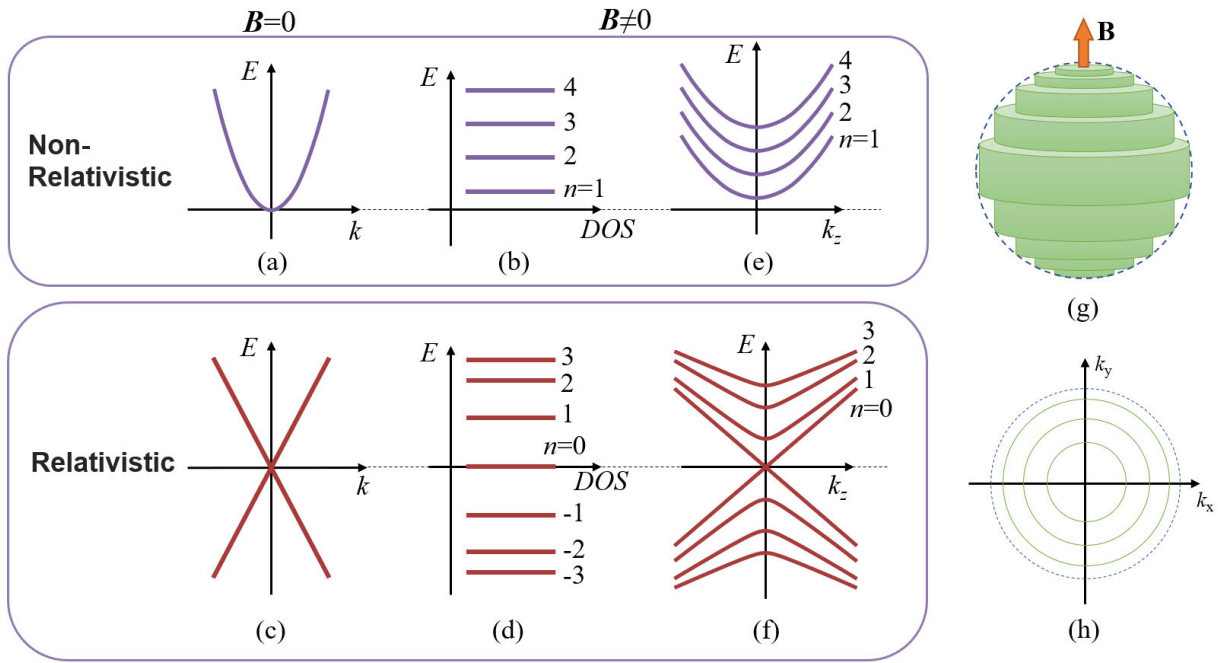

**Figure 3: Landau quantization.** (a) and (c): Schematics for energy – momentum dispersions of the (a) normal (non-relativistic) and (c) relativistic electrons. (b) and (d): Landau spectra for the 2D spinless (b) non-relativistic and (d) relativistic electrons. (e) and (f): Landau spectra for the 3D spinless (e) non-relativistic and (f) relativistic electrons with the magnetic field along $k_z$ direction ($\boldsymbol{B}//k_z$). (g) Landau tubes intersecting a 3D spherical Fermi surface. (h) Landau rings within the 2D Fermi surface (ring). (g) and (h) shows the scenario for non-relativistic electrons without the zeroth LL.

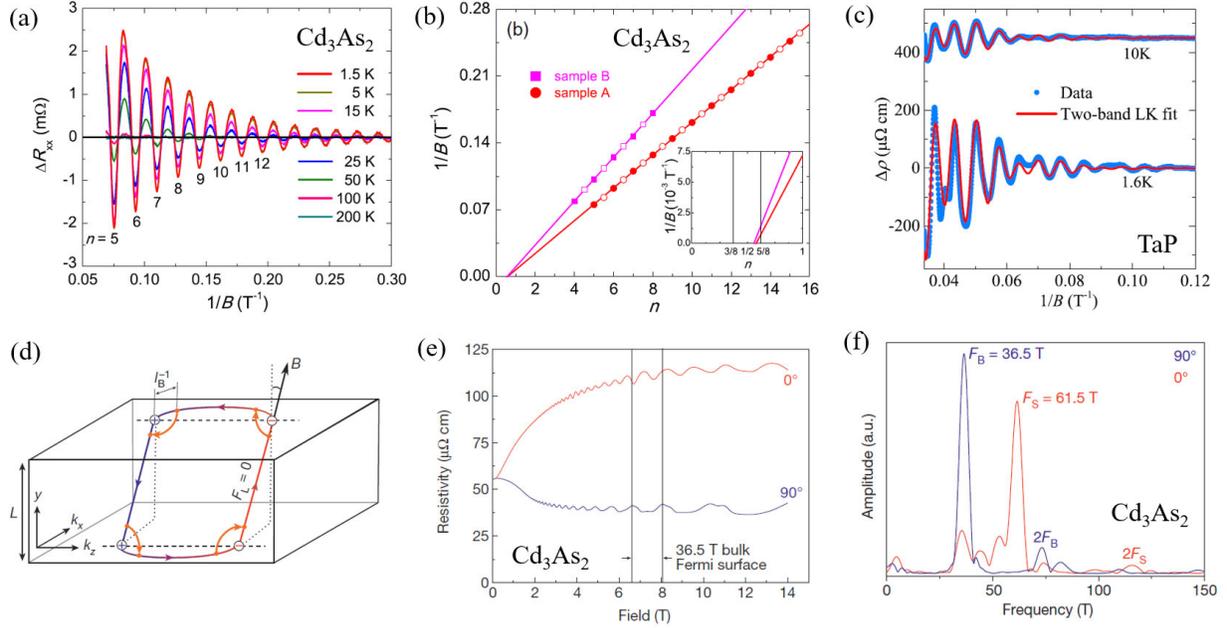

**Figure 4: Quantum oscillations in topological semimetals**. (a) The oscillatory component of resistance for $Cd_3As_2$, obtained via subtracting the smooth MR background, as a function of 1/B at various temperatures (178). (b) LL fan diagram constructed from SdH oscillations for two $Cd_3As_2$ samples. Inset: intercepts of the linear extrapolations of LL indices for the two samples. (178) (c) The oscillatory component of resistance for TaP, obtained via subtracting the smooth MR background, as a function of 1/B at various temperatures. The red solid lines show the fits of the oscillation data to the two-band LK model (192). (d) Mixed real and momentum space representation of the Weyl orbit, which consists of the Fermi arcs at the top and bottom surface connecting the projections of Weyl nodes with opposite chirality (labeled as "+" and "-", respectively), and the bulk states with fixed chirality (blue and red). (44) (e) and (f): Magnetoresistance at 2 K and its FFT for a thin (150 nm) slab sample, for magnetic field parallel (90°) and perpendicular (0°) to the surface. In addition to the bulk frequency $F_B$, another oscillation frequency corresponding to the surface state (FS) is observed for the perpendicular field. (44)

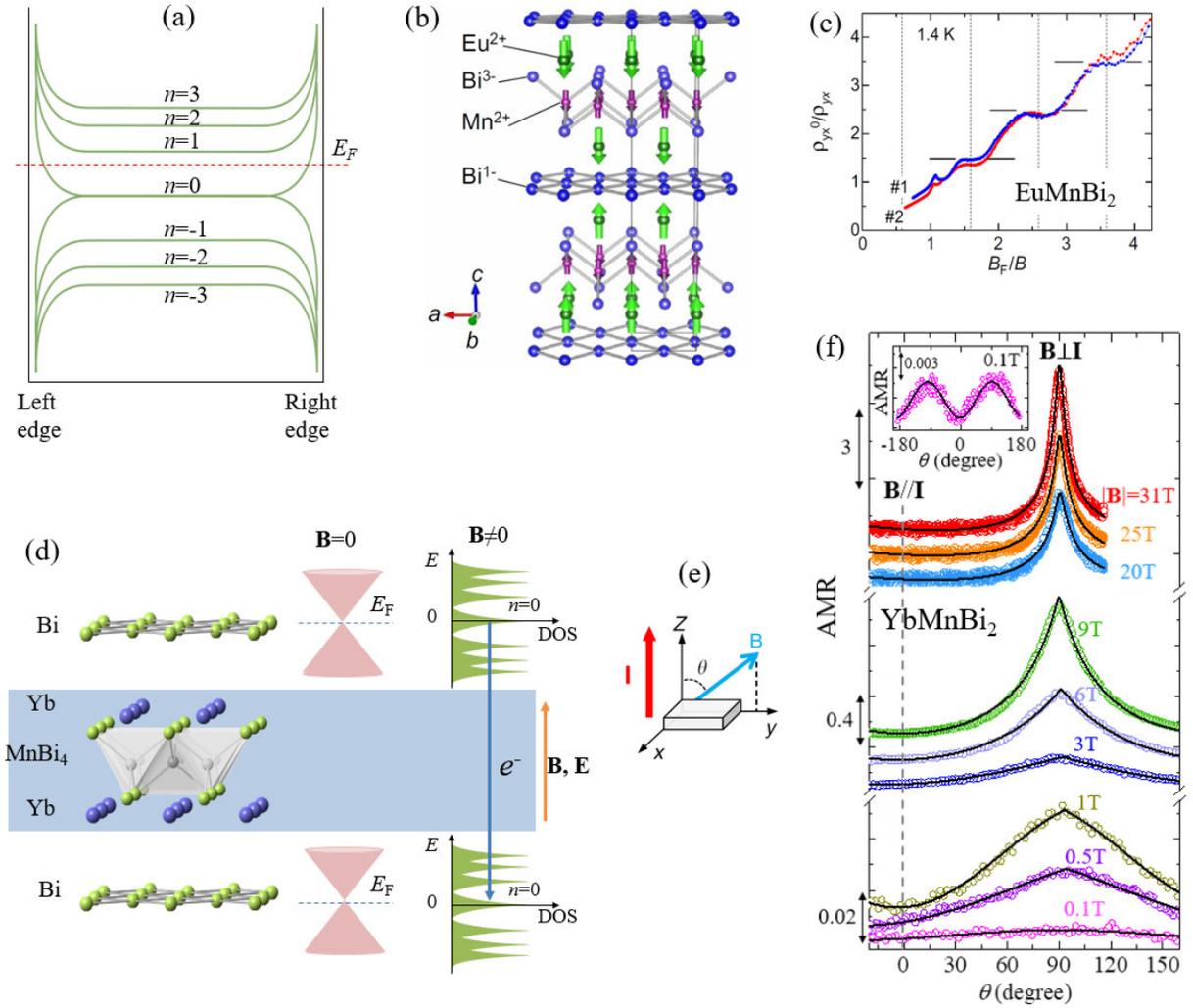

**Figure 5: Direct manifestations of the zeroth LL.** (a) Schematic of the real space Landau levels for relativistic electrons in a finite size 2D sample. (b) Crystal structure of EuMnBi$_2$ (177). (c) Normalized inverse Hall resistivity $\rho_{xy}^0/\rho_{xy}$ versus $B_F/B$ measured at 1.4 K for two EuMnBi$_2$ samples, where $B_F$ is the SdH oscillation frequency and $B = \mu_0(H + M)$ is the magnetic induction. (177) (d) Schematic of the interlayer tunneling of the zeroth LLs' relativistic fermions in YbMnBi$_2$. (247) (e) Experimental setup for the measurement of the angular dependence of interlayer magnetotransport. (247) (f) Angular dependence of interlayer resistance (AMR) measured under different fields up to 31T and at $T = 2$K, using the setup in (e). The black curves superimposed on the data represent the fits to the tunneling model. The inset shows the $\sin^2\theta$ dependence at low field. (247)

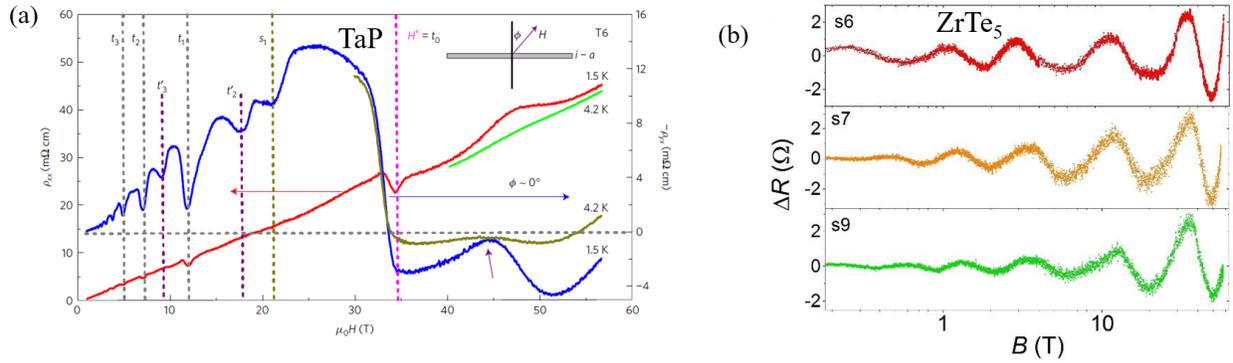

**Figure 6: Anomalous transport behavior beyond the quantum limit.** (a) Magnetic field dependence of the longitudinal ($\rho_{xx}$) and transverse ($\rho_{xy}$) resistivity at 1.5K and 4.2K for TaP. A steep drop with sign reversal for $\rho_{xy}$ are seen at high field. (273) (b) The oscillatory component of resistance $\Delta R$ at 4.2K of three ZrTe$_5$ samples (s5, s7, and s9) with log($B$) period. (274)

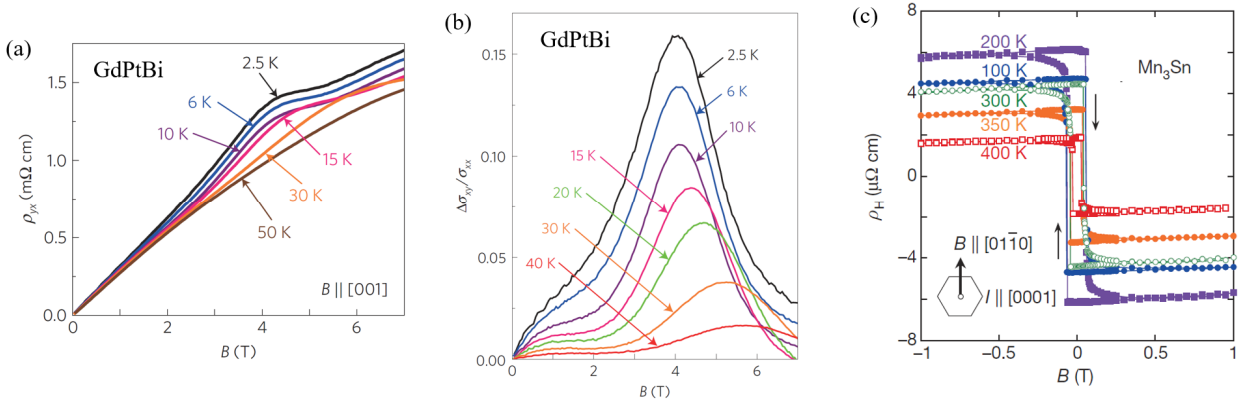

**Figure 7: Anomalous Hall effect.** (a) Magnetic field dependence of the transverse (Hall) resistivity $\rho_{xy}$ for GdPtBi, with field along the [001] direction. (298) (b) Anomalous Hall angle $\Delta\sigma_{xy}/\sigma_{xx}$ at different temperatures for GdPtBi. (298) (c) Magnetic field dependence of the Hall resistivity $\rho_H$ for Mn$_3$Sn. (100)

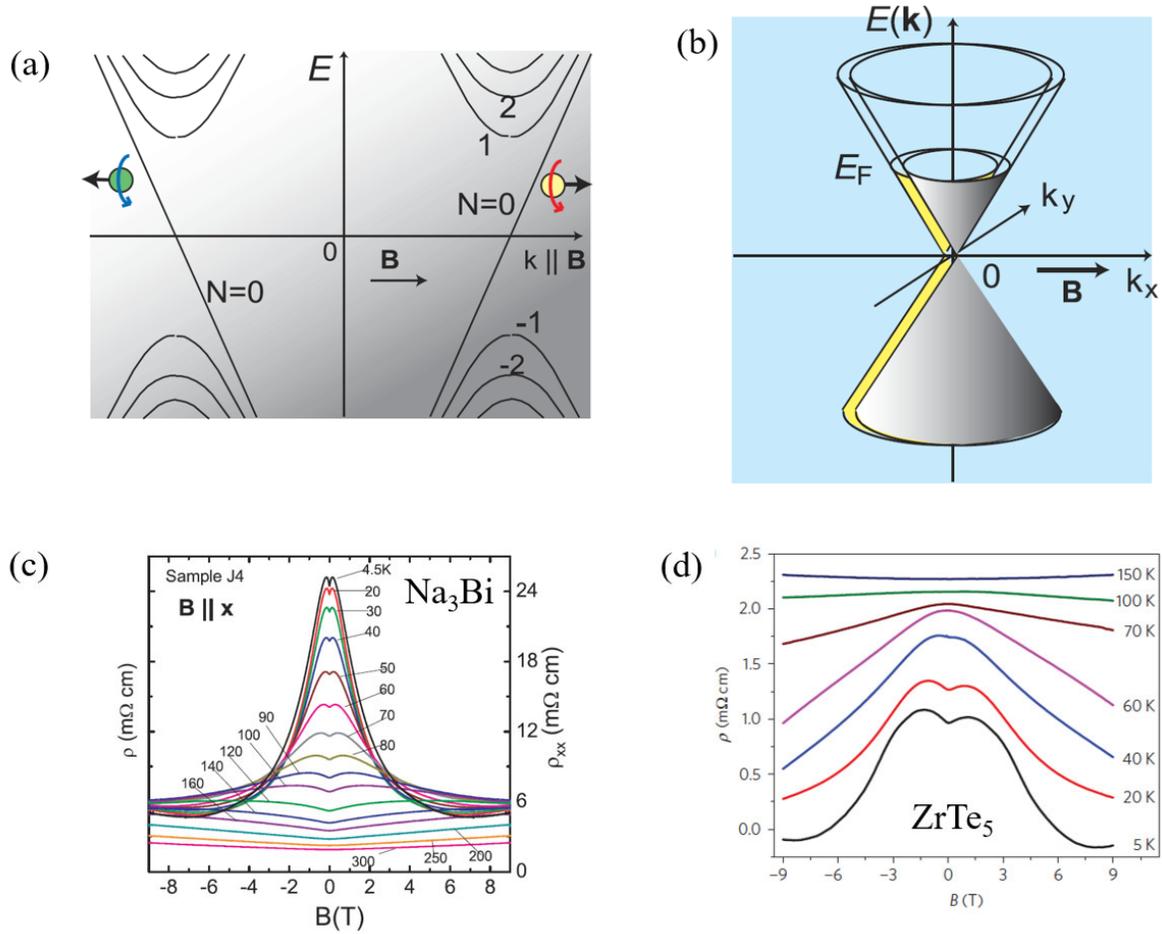

**Figure 8: Chiral anomaly and negative longitudinal MR.** (a) Schematic of chiral charge pumping between two Weyl cones with opposite chiralities under parallel magnetic and electric field. (105) (b) Magnetic field-induced Weyl state by lifting the spin degeneracy of a Dirac cone due to the Zeeman effect. (105) (c) Longitudinal $\rho_{xx}$ at various temperatures for $Na_3Bi$. Negative longitudinal MR is observed at lower temperatures. (105) (d) Longitudinal $\rho_{xx}$ at various temperatures for $ZrTe_5$. Negative longitudinal MR is observed at lower temperatures. (107)

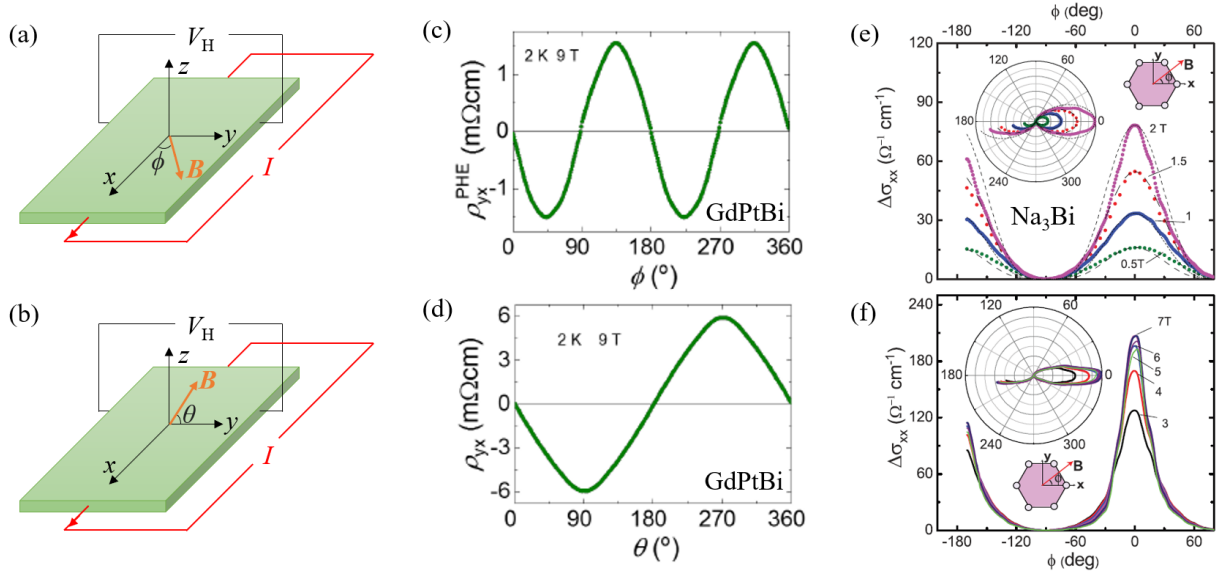

**Figure 9: Planar Hall effect (PHE) and AMR narrowing.** (a) Experimental setup for PHE. The magnetic field is rotated within the sample plane (the *x-y* plane). (b) Experimental setup for conventional Hall effect. The magnetic field is rotated from the out-of-plane direction toward the sample plane (the *y-z* plane). (c-d) Angular dependence of the (c) planar ($\rho_{xy}^{PHE}$) and (d) conventional ($\rho_{xy}$) Hall resistivity in GdPtBi at 9T and 2K, using the setup showing in (a) and (b), respectively. A two-fold symmetry is observed for the PHE, in contrast with the one-fold symmetry for the conventional Hall effect. (313) (e-f) Magnetic field orientation dependence of the magnetoconducivity [$\Delta\sigma_{xx} = \sigma_{xx}(B,\phi) - \sigma_{xx}(B,90º)$] for Na$_3$Bi at 4.5 K, measured at (a) low and (b) high magnetic fields. The insets show the same data in polar representation. The peak profiles in the angular dependence is clearly narrowed at high fields. (105)

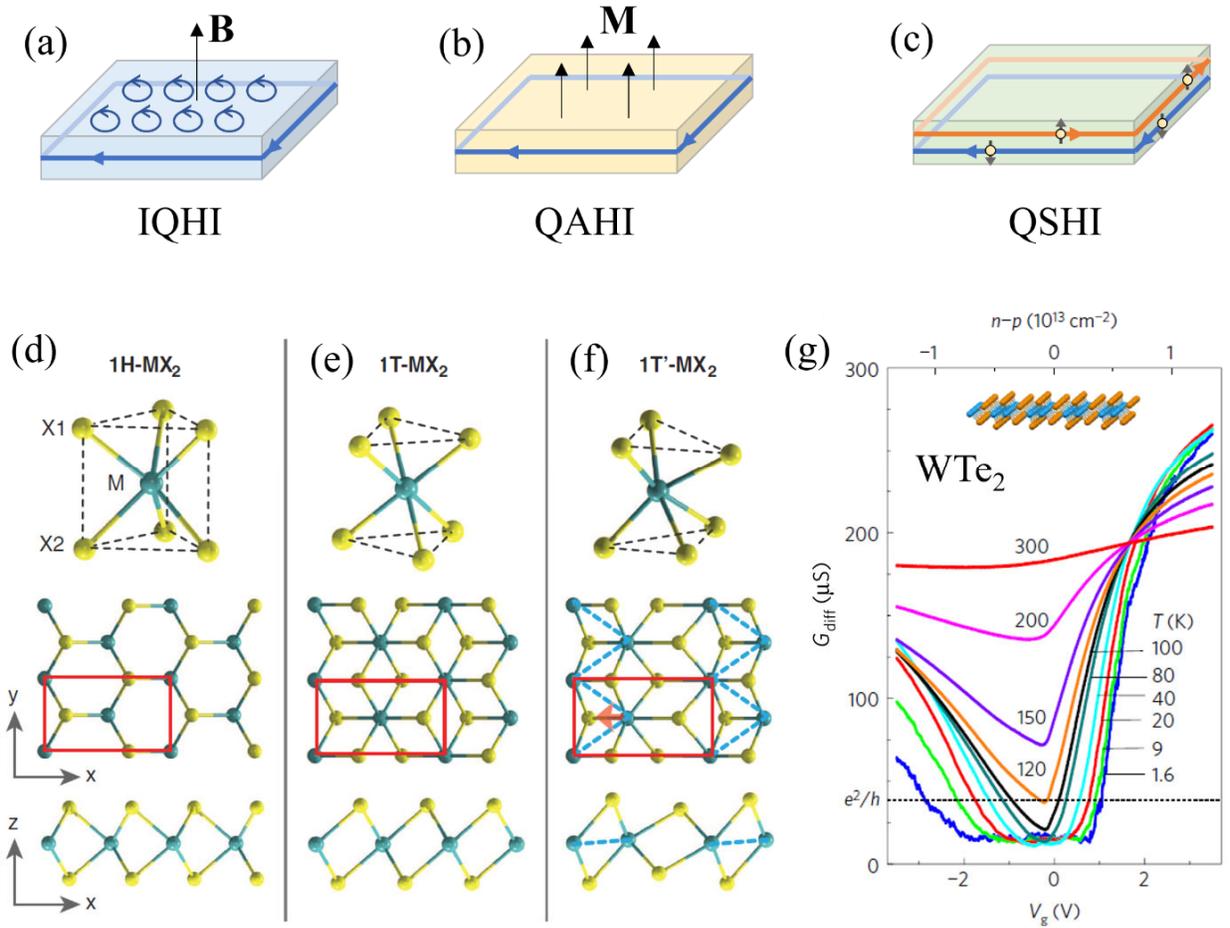

**Figure 10: Quantum Hall effects in various topological phases.** (a-c): Schematic for (a) the integer quantum Hall insulator (IQHI), (b) the quantum anomalous Hall insulator (QAHI), and (c) the quantum spin Hall insulator (QSHI) states. (d-f): The (d) 1H, (e) 1T, and (f) 1T' structures of monolayer transition metal dichalcogenides. (72) (g) Gate voltage dependence of the differential conductance of the monolayer WTe$_2$ at difference temperatures. (74)

Table 1: Parameters obtained from transport and quantum oscillation experiments at based temperatures (1.5-5 K), including MR at 9T, residual resistivity $\rho_{res}$, transport mobility $\mu_T$, quantum relaxation time $\tau_q$, quantum mobility $\mu_q$, and effective mass ratio $m^*/m_0$

| | MR at 9T | $\rho_{res}$ ($\mu\Omega$ cm) | $\mu_T$ (cm² V⁻¹ s⁻¹) | $\tau_q$ (ps) | $\mu_q$ (cm² V⁻¹ s⁻¹) | $m^*/m_0$ | Ref. |
|---|---|---|---|---|---|---|---|
| Cd$_3$As$_2$ | 34.5 - 1336 | 0.032 - 46.5 | 4 × 10³ - 8.7 × 10⁶ | 0.03 - 0.21 | 4,700 – 6,000 | 0.023 - 0.26 | (45; 172; 178; 236; 237; 272) |
| Na$_3$Bi | 5.69-97.1 | 1.72 - 87 | 5,500 - 78,900 | 0.0816 | | 0.11 | (105; 176) |
| TaAs family | 3 – 30,000 | 0.63 – 1.9 | 18,000 – 10,000,000 | 0.038 - 1.1 | 32,000 | 0.021 - 0.68 | (46; 179-183; 191-194; 225; 227; 243) |
| WTe$_2$ | 4,000 – 25,000 | 0.39 – 1.9 | 24,000 – 176,000 | | | 0.41 – 0.46 | (48; 184; 186; 231) |
| MoTe$_2$ | 2,653 | 28 | 16,000 - 58,000 | | | 0.8 - 2.9 | (337-340) |
| PtSn$_4$ | 1,000 – 2,100 | | | | 14,257 - 15,809 | 0.05 - 0.36 | (341-343) |
| PtBi$_2$ | 12,000 | | | | | | (171) |
| Pt(Te/Se)$_2$ | A few tens | | 3,600 – 5,500 | | | 0.11 - 3.6 | (229; 344) |
| PdTe$_2$ | A few tens | | | 0.18 - 0.65 | 1,293 – 6,209 | 0.04 - 1.16 | (229; 248) |
| AMn(Sb/Bi)$_2$ A = Ca, Sr, Ba, Yb | 1 | | 1,500 3,400 | – | | | (141; 143; 173; 175; 213; 234; 235; 238; 239; 247; 283; 345; 346) |
| WHM * | 1.3 – 1,4000 | 0.052 | 2,000-28,000 | 0.025-0.35 | 209 -10,000 | 0.025 - 0.27 1.32** | (156; 222; 226; 228; 232; 233; 245; 347) |
| | | | | | | | |

* MR, effective mass, and quantum relaxation time strongly varies in different *WHM* materials, possibly due to the spin-orbit coupling gap that vary with the atomic number.

** mass enhancement at low temperatures (245)